\documentstyle[prd,aps,epsf]{revtex}
\begin{document}
\title{
{\tt hep-ph/9801297}\hfill{\small { FZJ-IKP(TH)-1998-01}}\\
\hfill{\small { LPT-98-01}}\\ 
\hfill{\small { TRI-PP-97-73}}\\[2cm]
The form factors of the nucleon at small momentum transfer}

\vspace{1cm} 

\author{V\'eronique Bernard$^a$\footnote{email: bernard@sbghp4.in2p3.fr}, 
Harold W. Fearing$^b$\footnote{email: fearing@triumf.ca}, 
Thomas R. Hemmert$^b$\footnote{email:
  hemmert@triumf.ca}\footnote{Address after February 1$^{st}$, 1998: FZ
  J\"ulich, IKP (Th), D-52425 J\"ulich.}
and Ulf-G. Mei\ss ner$^c$\footnote{email: Ulf-G.Meissner@fz-juelich.de}}

\vspace{0.5cm}

\address{$^a$ Laboratoire de Physique Th\'eorique, BP 28, F-67037
 Strasbourg Cedex 2, France}
\address{$^b$ Theory Division, TRIUMF, 4004 Wesbrook Mall, Vancouver,
 BC, Canada V6T 2A3}
\address{$^c$ Forschungszentrum J\"ulich, Institut f\"ur Kernphysik (Th), D-52425 
 J\"ulich, Germany }
\maketitle

\thispagestyle{empty}

\vspace{2cm}

\begin{abstract}
We study the low energy expansion of the nucleon's electroweak form
factors in the framework of an effective chiral Lagrangian including pions,
nucleons and the $\Delta (1232)$. We work to third order in the
so--called small scale expansion and compare the results with the ones
previously obtained in the chiral expansion. In addition, these
calculations serve as a first exploratory study of renormalization
and decoupling within the small scale expansion.
\end{abstract}
\newpage

\section{Introduction and Summary}
The electromagnetic and axial structure of the nucleon as revealed
e.g. in elastic electron--nucleon and (anti)neutrino--nucleon scattering
as well as in charged pion electroproduction is parameterized in terms of the six
form factors $F_{1,2}^{p,n} (q^2)$ and $G_{A,P} (q^2)$
 (with $q^2$ the squared momentum transfer). The 
understanding of these form factors is of utmost importance in any theory or 
model of the strong interactions. At low energies, these form factors
can be Taylor expanded in $q^2$, with the terms linear in $q^2$ giving the
respective charge, magnetic and axial radii. It should not come as a
surprise that these different probes, i.e. the
isoscalar/isovector vector and the isovector axial current, see
different nucleon sizes. While the  electromagnetic radii have
a typical size of about 0.85~fm, the axial radius is significantly smaller,
about 0.65~fm. The pattern of these scales is related to the various
intermediate states dominating the response of an isoscalar/isovector
photon or the external axial current (mediated by the $W/Z$--bosons)
(for a very instructive model, see~\cite{UGM}). The current situation
concerning the theoretical understanding and experimental knowledge of
the electromagnetic form factors is reviewed in~\cite{klein}\cite{ugmf}\cite{dd}.

Chiral perturbation theory (ChPT) can be used to investigate the Green
functions of QCD quark currents in the low energy domain. The 
abovementioned form factors have been studied over the years in
the relativistic formulation of baryon ChPT~\cite{GSS} and in the
heavy baryon formulation (HBChPT), which admits a one--to--one
correspondence between the expansion in pion loops on one side 
and small momenta and quark (pion) masses on the other~\cite{BKKM}.
Here small means with respect to a typical hadronic scale, say the
proton mass.
In addition, the spectral functions of the isoscalar electromagnetic
as well as the isovector axial current have been worked out to 
two loop accuracy \cite{BKMFF}. This is mandated by the fact that in a dispersive
representation, the pertinent absorptive parts start to contribute
with the three--pion intermediate state and are therefore not
accessible in a one--loop calculation. It could also be shown
that the strong unitarity correction on the left wing of the
$\rho$--resonance in the isovector electromagnetic channel can be reproduced by
a one loop calculation~\cite{GSS}\cite{BKMFF}. Furthermore, the
relation between the axial radius extracted from (anti)neutrino--proton scattering
or pion electroproduction data was clarified in~\cite{BKMAX} and
precise predictions for the induced pseudoscalar coupling constant $g_P$
were made in Refs.\cite{BKMGP}\cite{FLMS}.

It has been argued for some time that for baryon ChPT, one has to include
the spin--3/2 decuplet (in SU(3)) or the $\Delta (1232)$ (in
SU(2))~\cite{jmd}.\footnote{From now on, we will only be concerned
  with the two flavor case.} This is motivated by the close proximity of this
nucleon resonance to the ground state and its very strong coupling to
the pion--nucleon--photon system. In fact, for QCD in the limit of
infinitively many colors, the nucleon and the delta resonance 
become degenerate in mass.
Therefore, in Refs.\cite{hhkplb}\cite{hhkbig} the so--called small
scale expansion was developed. Besides the two small parameters known
from the chiral expansion, the $N\Delta$ mass splitting $\Delta$ is introduced
as an additional small parameter and the methods of \cite{BKKM} have
been generalized to deal with such a situation. These three expansion
parameters are collectively called $\epsilon$. To assess
the accuracy of this novel approach and compare it to the conventional
chiral expansion, one has to systematically calculate observables and
compare the resulting predictions. Here, we concentrate on the
aforementioned form factors for essentially two reasons.
First,  these are relatively simple three--point functions and they
are known empirically to a good precision. Second, this work contains
the first exploratory study of ${\cal O}(\epsilon^3)$ 
renormalization in the small scale expansion. This latter topic is
only in its infant stage and many more details will have to be worked
out.

\bigskip

The pertinent results of this investigation can be summarized as
follows:

\begin{enumerate}

\item[1)] We have constructed the complete $\pi N \Delta \gamma$
  effective Lagrangian necessary to investigate the nucleons
  electroweak form factors in the small scale expansion to order
  $\epsilon^3$. In particular, we have performed the necessary 
  renormalization of the nucleon mass and various coupling constants,
  cf. Table~1. We have also discussed the decoupling of the delta
  degrees of freedom and how it affects the finite values of certain
  new counterterms related to the new scale $\Delta$.

\item[2)] We have considered in detail the isovector nucleon form
  factors at small momentum transfer. We recover the well--known
  result that the radii corresponding to the Dirac and Pauli form
  factor  explode in the chiral limit like $\log m_\pi$ and $1/m_\pi$,
  respectively. The new delta contributions bring the prediction for the Pauli 
  radius $r_2^v$ closer to the empirical value and lead to an improved
  description of the Pauli form factor for momentum transfer $|q^2|
  \le 0.2\,$GeV$^2$, cf. Fig.~2. The Dirac form factor 
  as given by the chiral and the small scale expansion agrees nicely
  with result based on dispersion relations.

\item[3)] The results for the isoscalar electromagnetic form factors
   differ from the ones in the chiral expansion only by a finite
   counterterm, which has no physical consequence.
   Combining these with the isovector form
   factors, the corresponding electric and magnetic proton form
   factors, as well as the magnetic neutron one, come rather close to the
   empirical dipole fit. The prediction for the neutron charge
   form factor is somewhat smaller than the best available
   parameterizations but consistent with the rather uncertain data
   (below $|q^2| \le 0.1\,$GeV$^2$), see figs.3a,b.

\item[4)] The axial form factor $G_A(q^2)$ receives only polynomial terms 
   up-to-and-including order $q^2$ in the momentum expansion.
   As in the case of the chiral expansion, the axial radius is
   determined by a finite low--energy constant. Delta effects
   only appear in the renormalization of the axial--vector coupling 
   constant $g_A \equiv G_A(0)$ but do not affect the $q^2$ dependence of this
   form factor to the order we are working. 

\item[5)] The prediction for the induced pseudoscalar form factor
  $G_P(q^2)$ and the coupling constant $g_P$ as well as the so--called
  Goldberger--Treiman discrepancy are also found not to be affected 
  by the delta to leading order in the small scale expansion, once the
  renormalization of the coupling constants has been accounted for
  (see table~2).
  $\Delta (1232)$ effects in the individual weak and electromagnetic form
  factors can therefore not explain the recently
  published TRIUMF value for $g_P$ obtained from a measurement of radiative
  muon capture \cite{Triumf}. Other $\Delta(1232)$ effects in radiative muon
  capture are also apparently small \cite{Beder}. 

\end{enumerate}

\section{HBChPT and the Small Scale Expansion}

\subsection{Effective Lagrangian, renormalization and decoupling}

In this section, we give the effective Lagrangian of the
pion--nucleon--$\Delta (1232)$--photon system necessary to work out
the nucleon's electroweak form factors at low energies. The chiral
counting is supplemented by an additional small parameter, the
$\Delta(1232)$--nucleon mass splitting $M_\Delta - M_N$, denoted as
$\Delta$. In contrast to the external momenta ($p$) and the pion mass
($m_\pi$), this
quantity does not vanish in the chiral limit of QCD. However, for a
very large number of colors, the nucleon and the $\Delta (1232)$ become
degenerate in mass. Furthermore, the $\Delta$--resonance is very strongly
coupled to the $\pi N \gamma$ system. One therefore considers an
expansion in a small parameter $\epsilon$, which in fact is a triple
expansion, since
\begin{equation}
\epsilon \in \{ p, m_\pi, \Delta \} \,\, .
\end{equation}
All these are small quantities compared to the typical hadronic scale
of about 1~GeV. This extends the chiral expansion in a logical fashion
for such  resonance degrees of freedom. It should, however, be kept in mind
that we are not dealing with a low--energy expansion in the strict sense,
where all expansion parameters vanish in a certain limit. For more
details, we refer the reader to~\cite{hhkplb},\cite{hhkbig}. 
 
Consider first the lowest order Lagrangians of dimension one and two.
These are special since loops only start at dimension three and thus
the pertinent coupling constants are not affected by loop effects to this order and
are therefore finite.
Denoting by $N$ the (heavy) nucleon isodoublet field and by
$T_\lambda^i$ the Rarita--Schwinger representation of the (heavy) spin--3/2
field (the delta), we have 
\begin{eqnarray}
L_{\pi N}^{(1)}&=& \bar{N} (i v \cdot D + \dot{g}_A S\cdot u) N~, \\ 
L_{\pi \Delta}^{(1)}&=& -\bar{T}_\mu^i \bigl( i v\cdot D^{ij} - \Delta_0
\xi_{3/2}^{ij} + \dot{g}_1 S \cdot u^{ij} \bigr) g^{\mu\nu} T_\nu^j~,\\ 
L_{\pi N\Delta}^{(1)}&=& \dot{g}_{\pi N \Delta}\bar{T}_\mu^i w^\mu_i N
+ {\rm h.c.}~,\\ 
L_{\pi N}^{(2)}&=& 
{\bar {N}}\biggl\{ \frac{1}{2 M_0}  (v\cdot
D)^2-\frac{1}{2 M_0}{D \cdot D} 
-\frac{i\, \dot{g}_A}{2 M_0}\{ S\cdot D , v \cdot u \} 
 + c_1 \,{\rm Tr}(\chi_+) \nonumber \\
& &\, -\frac{i}{4 M_0}[S^\mu , S^\nu ]\, 
\biggl[(1+\dot{\kappa}_v) \, f_{\mu\nu}^+ 
+ 2(1+ \dot{\kappa}_s)\,v_{\mu\nu}^{(s)} \biggr] +\ldots
\biggr\} {N} \,\, .
\end{eqnarray}
The ellipsis denotes other terms not relevant for the further
discussion. Here, $\dot Q$ denotes that a quantity $Q$ is taken in 
the chiral SU(2) limit,\footnote{The exception to this are the
 nucleon mass, the $N\Delta$--splitting and the pion decay constant in
 the chiral limit, which are denoted by $M_0$, $\Delta_0$ and $F_0$,
 respectively. To the accuracy we are working, we can set
 $\Delta_0 = \Delta$.} $m_u = m_d =0$ and $m_s$ fixed, i.e.
\begin{equation}
Q = \dot Q \, [ 1 + {\cal O}(m_{u,d}^\alpha)] \,\, ,
\end{equation} 
with the power $\alpha = 1/2$ or $1$. Furthermore, $i,j=1,2,3$ are isospin
indices, $\xi_{3/2}^{ij}$ is an isospin--3/2 projection operator, 
$D_{\mu}^{ij}$ is the covariant derivative for spin 3/2, isospin 3/2 systems
~\cite{hhkplb,hhkbig} and
$S_\mu$ is the Pauli--Lubanski spin vector. We also 
employ
\begin{eqnarray}
w_\mu^i &=& \frac{1}{2} {\rm Tr}\left(\tau^i u_\mu\right) \; , \\
u_\mu^{ij} &=& \xi^{ik}_{3/2} u_\mu \xi^{kj}_{3/2} \; . 
\end{eqnarray}
The scalar source $\chi_+$
includes the explicit chiral symmetry breaking through the pion mass
$m_\pi$ and $\kappa_{s,v}$ are the isoscalar/isovector anomalous
magnetic moments of the nucleon.
We follow here  the notation of \cite{BKM97}, i.e. the isovector
component of the photon is encoded in $f_{\mu\nu}^+$
whereas the isoscalar sources are contained in $v_{\mu\nu}^{(s)}$.
For more details, we refer the reader to 
Refs.\cite{BKM97}\cite{BKMREV} (nucleon sector) and 
Refs.\cite{hhkplb}\cite{hhkbig} ($N\Delta$ sector).
Throughout, we  use $g_{\pi N \Delta} = 1.05$, $g_A=1.26$, $M_N=0.938\,$GeV,
$m_\pi=0.14\,$GeV, $F_\pi=0.0925\,$GeV and $\Delta=0.294\,$GeV.
The $\Delta\Delta \pi$ coupling constant $g_1$ 
appears only in the renormalization of $g_A$
(cf. Table 2) and we therefore do not need to specify its value.

At order $\epsilon^3$, there are two types of contributions. First,
there are $1/M_N^2$ corrections with fixed coefficients. These can
be deduced in the standard fashion from the lagrangian,
\begin{eqnarray}
L_{\pi N}^{(3), {\rm fixed}}&=& \bar{N} \left[ \gamma_0 \tilde{{\cal
      B}}^\dagger_N \gamma_0  \tilde{{\cal C}}^{-1}_N  \tilde{{\cal
      B}}_N + \gamma_0 {\cal B}^\dagger_{\Delta N} \gamma_0 {\cal
      C}_\Delta^{-1} {\cal B}_{\Delta N}\right] N \,\, ,
\end{eqnarray}
with the explicit form of the various matrices $\tilde{{\cal B}}_N, {\cal B}_{\Delta
  N}, \ldots$ given in~\cite{hhkbig}. Second,
the ${\cal O}(\epsilon^3)$ counterterms are given by the general structure
(the explicit form of the operators is given in the later sections)
\begin{eqnarray}
L_{\pi N}^{(3)}&=&\frac{1}{(4\pi F_\pi)^2}\sum_{i=1}^{22}B_i \bar{N} 
                  O_{i}^{(3)} N + \frac{1}{(4\pi F_\pi)^2} \sum_{j=1}^{9} 
                  \tilde{B}_j \bar{N} O_{j}^{(3)} N \nonumber \\
               &+&\frac{\Delta}{(4\pi F_\pi)^2}\sum_{i=23}^{29}B_i \bar{N} 
                  O_{i}^{(2)} N +
                  \frac{\Delta^2}{(4\pi F_\pi)^2}\sum_{i=30}^{31}B_i \bar{N} 
                  O_{i}^{(1)} N +
                  \frac{\Delta^3}{(4\pi F_\pi)^2}\;B_{32}\bar{N} 
                  O_{32}^{(0)} N \; , \label{eq:l3}
\end{eqnarray}
with
\begin{eqnarray}
B_i &=&B_{i}^r(\lambda)+\beta_i 16\pi^2 L \,\, ,\nonumber \\
L   &=& \frac{\lambda^{d-4}}{16 \pi^2} \biggl[ \frac{1}{d-4} +
\frac{1}{2} (\gamma_E - 1 - \ln 4\pi ) \biggr] \,\, ,
\end{eqnarray}
with $\gamma_E$ the Euler--Mascheroni constant
and $\lambda$ the scale of dimensional regularization. The
$\beta$--functions $\beta_i$ ($i=1,22$)
for the nucleon sector are tabulated in \cite{Ecker}.
The additional finite terms $\tilde{B}_j$ are scale--independent and
can be found in \cite{fms}. 
We note that in principle there 
are ten\footnote{Our analysis will suggest that $B_{29}$ is finite, see Table
1.} new counterterms \cite{hhkbig} in Eq.(\ref{eq:l3}) in addition 
to the 22 counterterms of ${\cal O}(p^3)$ HBChPT \cite{Ecker} and the 9 
finite terms at ${\cal O}(p^3)$ \cite{fms}\footnote{\label{fn1} 
We note that by construction the 
${\cal O}(\epsilon^n)$ ``small scale expansion'' has been set up that 
one can recover both the ${\cal O}(p^n)$ HBChPT lagrangians and the 
${\cal O}(p^n)$ results in the ``limit'' where all $\Delta$(1232) dependent
couplings are set to zero.}. However, the 10 new counterterms
in the nucleon sector are unobservable in the sense that they do not bring 
about any new vertices for the theory as they 
correspond to lower order HBChPT Lagrangians multiplied with the new scale 
$\Delta$. They will therefore always enter in combinations with $\Delta$ 
independent couplings or counterterms (see Table 2) and we could choose their 
finite parts $B_{i}^r(\lambda)|_{i=23...32}$ in order to guarantee decoupling 
of the delta-resonance in the limit $\Delta\rightarrow\infty$ without  
affecting measurable quantities. Nevertheless, throughout this work we will 
choose to write all
results in terms of physical coupling constants and therefore completely 
avoid the question of the finite sizes for these terms. Once our results
are written in terms of physical quantities the decoupling of $\Delta$(1232)
contributions in the limit $\Delta \to \infty$ must of course be 
``automatically'' guaranteed. For a general 
discussion of decoupling in the effective meson--baryon field theory we
refer to \cite{gaze,ulfn*} and will
come back to this point later. In Table~1 we have collected the counterterm
structures which are of relevance for the problem under
consideration. We give the pertinent operators together with their
$\beta$ functions in the chiral and in the small scale expansion, to
order $p^3$ and $\epsilon^3$, respectively, and indicate in which
quantity they appear. 

\subsection{Nucleon mass and wave function renormalization}

We only need the leading term in the renormalization of the nucleon mass
which is not modified by delta degrees of freedom:
\begin{equation}
M_N=M_0 - 4 c_1 m_{\pi}^2 + {\cal O}(\epsilon^3) \,\, ,
\end{equation}
where the LEC $c_1$ can e.g. be extracted from the pion--nucleon $\sigma$--term.
Note that at order $\epsilon^3$, the dimension zero 
operator $O^{(0)}_{32}$ leads to a finite
shift of the nucleon mass in the chiral limit. 
For the deltas treated relativistically, this shift is
known to be infinite \cite{bkmz}. For a more detailed study
of the nucleon mass in the small scale expansion we refer to the talk by
Kambor~\cite{jkmz}.
The nucleon wave function renormalization constant $Z_N$ to 
${\cal O}(\epsilon^3)$ reads
\begin{eqnarray}
Z_{N}^{(3)} &=&1-\frac{1}{(4\pi F_{\pi})^2}\left\{ \left(\frac{3}{2}g_{A}^2
               +4 g_{\pi N\Delta}^2\right) m_{\pi}^2 -16 g_{\pi N\Delta}^2 
               \Delta \sqrt{\Delta^2 -m_{\pi}^2}\;\log R  \right.
               \nonumber \\
            & &\phantom{1-\frac{1}{(4\pi F_{\pi})^2} } \left.
               +\;8 m_{\pi}^2 B_{20}^r(\lambda)+\left(\frac{9}{2}g_{A}^2 
               m_{\pi}^2+8 g_{\pi N\Delta}^2 m_{\pi}^2 \right) \log\left[ 
               \frac{m_\pi}{\lambda}\right] +\Delta^2 B_{30}^r(\lambda)
               -16 g_{\pi N\Delta}^2 \Delta^2 
               \log\left[\frac{m_\pi}{\lambda}\right] \right\} \; ,
\end{eqnarray}
with 
\begin{equation}
R=\frac{\Delta}{m_\pi}+\sqrt{\frac{\Delta^2}{m_{\pi}^2}-1} = 3.95~.
\end{equation}
In this and most subsequent formulas, to the order required, all couplings,
mass terms and decay constants have been expressed in terms of their {\it 
physical} values as detailed in Table~2.
We note that some authors employ additional momentum-dependent terms in $Z_N$
(see e.g. the discussion in refs. \cite{FLMS,EM97}), which we have
chosen to be
absorbed in the relativistic normalization of our nucleon spinors. Physical 
results, of course, do not depend on this choice, as the nucleon $Z$--factor 
is not a measurable quantity but only needed to perform the necessary 
renormalizations.

\bigskip

\section{Isovector Vector Form Factors}
The structure of the nucleon probed with virtual photons is encoded in
two form factors, called $F_{1,2} (q^2)$, with $q^2$ the squared
four--momentum transfer. Since the photon itself has an isoscalar and an
isovector component, it is natural to decompose these form factors
into corresponding isoscalar/isovector ($s/v$) parts. In this section,
we are concerned with the isovector form factors. To one loop, these
encode more information than their isoscalar counterparts because the
pertinent spectral functions start with the two--pion cut and thus
lead to a non--polynomial contribution at this order.

\subsection{Definition}
Consider the nucleon matrix element of the isovector component of the
quark vector current $V_\mu^i =\bar q \gamma_\mu (\tau^i /2) q$, 
which involves a vector and a tensor form factor,
\begin{eqnarray}
\langle N(p_2)|V_{\mu}^i(0)|N(p_1)\rangle=\bar{u}(p_2)\left[F_{1}^{\;v}(q^2)\;
\gamma_\mu+\frac{i}{2M_N}F_{2}^{\;v}(q^2)\;\sigma_{\mu\nu}
q^\nu\right]u(p_1) \times \eta^\dagger\frac{\tau^i}{2}\eta \,\, ,
\end{eqnarray}
where $u(p)$ is a Dirac spinor with isospin component $\eta$ and
$q^2 = (p_2 - p_1)^2$ is the invariant momentum transfer squared.
$F_1 (q^2)$ and $F_2(q^2)$ are the Dirac and the Pauli form factor,
respectively, subject to the normalizations
\begin{equation}
F_1^v (0) =1 \,\, , \quad F_2^v (0) = \kappa_v \,\, ,
\end{equation}
with  $\kappa_v=3.71$ the isovector nucleon anomalous magnetic moment.
As pointed out in \cite{BKKM}, in the heavy baryon approach it is most
natural to work in the brick--wall (Breit) frame since the nucleons
(and also the deltas) are essentially heavy static
sources. Furthermore, the Breit frame allows for a unique translation
of the Lorentz--covariant matrix elements into non--relativistic ones.
Thus performing the non-relativistic reduction in the Breit frame
and utilizing the heavy-mass decomposition
\begin{eqnarray}
p_{1}^\mu \rightarrow M_0 v^\mu +r_{1}^\mu, \quad p_{2}^\mu \rightarrow M_0 
v^\mu +r_{2}^\mu,
\end{eqnarray}
one obtains 
\begin{eqnarray}
\langle N(p_2)|V_{\mu}^i(0)|N(p_1)\rangle=\frac{1}{N_1 N_2}
\bar{u}_v(r_2)\left[\tilde{G}_{E}^{\;v}(q^2)\;v_\mu
+\frac{1}{M_N}\tilde{G}_{M}^{\;v}(q^2)\left[
S_\mu,S_\nu\right] q^\nu\right]u_v(r_1) \times \eta^\dagger\frac{\tau^i}{2}\eta
\; ,
\end{eqnarray}
with 
\begin{eqnarray}
u_v(r)&=&P_v^+ \; u(p) \; ,
         \quad P_v^+=\frac{1}{2}\left(1+\not{v}\right) \; ,
         \quad N_i = \sqrt{\frac{E_i + M_N}{2M_N}} \, \, , \, \, (i =1,2) \; . 
\end{eqnarray}
Here, $r_{1,2}^\mu$ are small (``soft'') momenta in the sense that
$r_{1,2}^\mu v_\mu \ll M_0$. For the specific choice $v^\mu=(1,0,0,0)$ of the
the velocity vector $v_\mu$ they read
\begin{eqnarray}
r_{1}^{\mu}=\left(E-M_0,-\frac{\vec{q}}{2}\right), \quad r_{2}^\mu=\left(E-
M_0,+\frac{\vec{q}}{2}\right), \quad q^\mu=\left(0,\vec{q} \, \right)
\end{eqnarray}
and the form factors $\tilde{G}_E(q^2),\tilde{G}_M(q^2)$ 
then correspond to the well-known
electric, $G_E(q^2)$, and magnetic, $G_M(q^2)$, Sachs form factors.
These are connected to the previously defined
Dirac/Pauli form factors via
\begin{eqnarray}
G_{E}^{\;v}(q^2)&=&F_{1}^{\;v}(q^2)+\frac{q^2}{4M_{N}^2}F_{2}^{\;v}(q^2)~, \\
G_{M}^{\;v}(q^2)&=&F_{1}^{\;v}(q^2)+F_{2}^{\;v}(q^2)~. \label{eq:def}
\end{eqnarray}
The Sachs form factors are normalized to the isovector electric charge
and the isovector total magnetic moment of the nucleon, respectively.
We now turn to the chiral expansion of these various isovector form
factors.

\subsection{Chiral Input}

In Fig.1, the pertinent one loop graphs for the ${\cal O}(\epsilon^3)$
calculation are shown. The following pieces of the dimension three
chiral Lagrangian are needed,
\begin{eqnarray}
{\cal L}_{\pi N}^{(3)}&=& \frac{1}{(4\pi F_\pi)^2} \bar{N} \left\{ B_{10} D^\mu
                          f_{\mu\nu}^+ v^\nu 
                          + B_{20} \left({\rm Tr} (\chi_+) i v \cdot D
                          + {\rm h.c.}
                          \right)
                          + B_{28} \Delta i \left[S^\mu,S^\nu\right] 
                          f_{\mu\nu}^+ 
                          + B_{30} \Delta^2 i v \cdot D \right\} N \nonumber \\
                      & & + \frac{1}{2 M_0} \; \bar{N} \left( \gamma_0 
                          B_{N}^{(2)\dagger}\gamma_0 B_{N}^{(1)} + \gamma_0 
                          B_{N}^{(1)\dagger}\gamma_0 B_{N}^{(2)}\right) N
                          - \frac{1}{(2 M_0)^2} \; \bar{N} \gamma_0 
                          B_{N}^{(1)\dagger}\gamma_0 \left(i v \cdot D
                          + \dot{g}_A S \cdot u \right) B_{N}^{(1)} N \; .
\end{eqnarray}
Here, $B_{10}, B_{20}, B_{28}, B_{30}$ are ${\cal O}(\epsilon^3)$ counterterms 
participating 
in the renormalization of the loop diagrams 1a...1f. When comparing with
the set of infinite counterterms of ${\cal O}(p^3)$ HBChPT \cite{Ecker} we note
that $B_{28}, B_{30}$ are new counterterms in the nucleon--sector due to the presence
of the additional dimensional scale $\Delta$. Note that $B_{20}$ and
$B_{30}$ only contribute to the nucleon $Z$--factor whereas $B_{10}$
and $B_{28}$ are related to the isovector charge radius and the isovector anomalous
magnetic moment, see below and Table~2. Finally, the matrices $B_{N}^{(i)}$ 
which govern the fixed $1/M^2$ corrections can be found in refs. 
\cite{BKKM,hhkbig}.

\subsection{Results}
Apart from the loop graphs shown in Fig.~1, one has Born term
contributions. Adding these, the electric
isovector form factor is given by (see app.~A for explicit expressions
of the various loop diagrams depicted in Fig.~1) 
\begin{eqnarray}
G_{E}^{\;v}(q^2)&=&1+\kappa_v \frac{q^2}{4 M_{N}^2} + \frac{1}{(4\pi F_\pi)^2}
                   \left\{ q^2 \left(\frac{68}{81} g_{\pi N\Delta}^2 - \frac{
                   2}{3}g_{A}^2-2 B_{10}^{(r)} \right) + q^2 \left(\frac{40}{
                   27} g_{\pi N\Delta}^2-\frac{5}{3}g_{A}^2-\frac{1}{3}\right)
                   \log\left[\frac{m_\pi}{\lambda}\right] \right. \nonumber \\
                & &\phantom{1+\kappa_v \frac{q^2}{4 M_{N}^2} + \frac{1}{
                   (4\pi F_\pi)^2} }
                   + \int_{0}^{1}dx \left[\frac{16}{3}\Delta^2 g_{
                   \pi N\Delta}^2+m_{\pi}^2 \left(3 g_{A}^2+1-\frac{8}{3}g_{
                   \pi N\Delta}^2 \right) \right. \nonumber \\
                & &\phantom{1+\kappa_v \frac{q^2}{4 M_{N}^2} + \frac{1}{
                   (4\pi F_\pi)^2} + \int_{0}^{1}dx \; } \left. 
                   -q^2 x(1-x)\left(5 g_{A}^2+1-\frac{
                   40}{9}g_{\pi N\Delta}^2\right)\right] \log\left[\frac{
                   \tilde{m}^2}{m_{\pi}^2}\right] \nonumber \\
                & &\phantom{1+\kappa_v \frac{q^2}{4 M_{N}^2} + \frac{1}{
                   (4\pi F_\pi)^2} }
                   + \int_{0}^{1}dx \left[ \frac{32}{9} g_{\pi N\Delta}^2 q^2 
                   x(1-x) \frac{\Delta \log R[ \tilde{m}^2 ]}{
                   \sqrt{\Delta^2-\tilde{m}^2}} \right]\nonumber \\
                & &\phantom{1+\kappa_v \frac{q^2}{4 M_{N}^2} + \frac{1}{
                   (4\pi F_\pi)^2} }\left.
                   - \int_{0}^{1}dx \; \frac{32}{3}g_{\pi N\Delta}^2 \Delta \left[\sqrt{
                   \Delta^2-m_{\pi}^2}\log R -\sqrt{\Delta^2-\tilde{m}^2}\log 
                   R[ \tilde{m}^2] \right] \right\} \; ,
\end{eqnarray}
with 
\begin{equation}
R\left[\tilde{m}^2\right]
= \frac{\Delta}{\tilde{m}}+\sqrt{\frac{\Delta^2}{\tilde{m}^2}-1}~, \quad
\tilde{m}^2 = m_\pi^2 - q^2 x (1-x)~.
\end{equation}
We note that for the isovector electric form factor, decoupling requires 
that in the
limit $\Delta \to \infty$ the whole delta contribution can be absorbed
in the value of $B^r_{10}$.

Analogously, the magnetic isovector form factor yields
\begin{eqnarray}
G_{M}^{\;v}(q^2)&=&1+\kappa_v - g_{A}^2 \frac{4\pi M_N}{(4\pi F_\pi)^2} 
                   \int_{0}^{1}dx\left\{ \sqrt{\tilde{m}^2}-m_{\pi}\right\}
                   \nonumber \\
                & &\phantom{1+\kappa_v} +\frac{32}{9}g_{\pi N\Delta}^2\frac{
                   M_N \Delta}{
                   (4\pi F_\pi)^2}\int_{0}^{1}dx \left\{ \frac{1}{2}\log\left[
                   \frac{\tilde{m}^2}{4\Delta^2}\right]-\log\left[\frac{m_\pi}{
                   2\Delta}\right] \right. \nonumber \\
                & &\phantom{1+\kappa_v +\frac{32}{9}g_{\pi N\Delta}^2\frac{
                   M_N \Delta}{(4\pi F_\pi)^2}\int_{0}^{1}dx} \left. 
                   +\frac{\sqrt{\Delta^2-\tilde{m}^2}}{\Delta}
                   \log R\left[\tilde{m}^2\right]
                   -\frac{\sqrt{\Delta^2-m_{\pi}^2}}{
                   \Delta}\log R\right\} \; ,
\end{eqnarray}
with the renormalization of $\kappa_v$ given in Table~2. Since the 
(renormalized) expression 
for $G_M^v (q^2)$ is free of counterterms,
decoupling of the $\Delta (1232)$ for the isovector magnetic form
factor demands that as $\Delta \to \infty$, the delta contributions
have to vanish.
In addition to the decoupling limit $\Delta \to \infty$ one can also recover 
the previously obtained ${\cal O}(p^3)$ HBChPT results for 
$G_{E}^v(q^2), G_{M}^v(q^2)$ of \cite{BKKM,FLMS} for the limit 
$g_{\pi N\Delta} \to 0$, as promised in footnote \ref{fn1}.
To make the comparison with the work of Fearing et al. one
must realize that the Ecker-Moj\v zi\v s ${\cal O}(p^3)$ HBChPT
Lagrangian \cite{EM96} has had some equation-of-motion (EOM) terms transformed 
away via field-redefinitions, which leads to slightly different expressions
for the beta-functions in table 1 and different expressions for the coupling 
constant renormalizations in table 2. Once one writes the results for the
form factors in terms of physical couplings and masses, all these apparent 
differences of course disappear.

Heavy mass methods like HBChPT or the small scale expansion ``naturally'' yield
non-relativistic results for the quantities of interest. It is therefore 
the Sachs form factors $G_E(q^2),G_M(q^2)$, which have a simple nonrelativistic
physical interpretation, that 
can be read off directly from the chiral amplitudes expressed in the Breit
frame. In order to obtain expressions for the relativistic form factors 
$F_{1}^{\;v}(q^2),\; F_{2}^{\;v}(q^2)$ one has to work a little bit.
One possible approach consists in casting the heavy baryon amplitudes 
back into a relativistic form, which was pursued in ~\cite{FLMS}. In this work
we follow a different path. Keeping in mind that the 
${\cal O}(\epsilon^3)$ calculation is sensitive at most to $1/M_{N}^2$ 
structures, one can read off the desired connection via Eq.(\ref{eq:def}):
\begin{eqnarray}
G_{E}^{\;v}(q^2)&=&F_{1}^{\;v}(q^2)+\frac{q^2}{4M_{N}^2}F_{2}^{\;v}(0) + 
                   {\cal O}(1/M_{N}^3) \\
\frac{1}{M_N}G_{M}^{\;v}(q^2)&=&\frac{1}{M_N}F_{1}^{\;v}(0)+\frac{1}{M_N}
                                F_{2}^{\;v}(q^2) + {\cal O}(1/M_{N}^3)
\end{eqnarray}
Therefore, the ${\cal O}(\epsilon^3)$ expressions of the Dirac and
Pauli form factors read
\begin{eqnarray}
F_{1}^{v}(q^2)&=& 1+\frac{1}{(4\pi F_\pi)^2}
                   \left\{ q^2 \left(\frac{68}{81} g_{\pi N\Delta}^2 - \frac{
                   2}{3}g_{A}^2-2 B_{10}^{(r)} \right) + q^2 \left(\frac{40}{
                   27} g_{\pi N\Delta}^2-\frac{5}{3}g_{A}^2-\frac{1}{3}\right)
                   \log\left[\frac{m_\pi}{\lambda}\right] \right. \nonumber \\
                & &\phantom{1+\frac{1}{(4\pi F_\pi)^2} }
                   + \int_{0}^{1}dx \left[\frac{16}{3}\Delta^2 g_{
                   \pi N\Delta}^2+m_{\pi}^2 \left(3 g_{A}^2+1-\frac{8}{3}g_{
                   \pi N\Delta}^2 \right) \right. \nonumber \\
                & &\phantom{1+\frac{1}{(4\pi F_\pi)^2} + \int_{0}^{1}dx \; 
                   } \left. 
                   -q^2 x(1-x)\left(5 g_{A}^2+1-\frac{
                   40}{9}g_{\pi N\Delta}^2\right)\right] \log\left[\frac{
                   \tilde{m}^2}{m_{\pi}^2}\right] \nonumber \\
                & &\phantom{1+\frac{1}{(4\pi F_\pi)^2} }
                   + \int_{0}^{1}dx \left[ \frac{32}{9} g_{\pi N\Delta}^2 q^2 
                   x(1-x) \frac{\Delta \log R[ \tilde{m}^2 ]}{
                   \sqrt{\Delta^2-\tilde{m}^2}} \right]\nonumber \\
                & &\phantom{1+\frac{1}{(4\pi F_\pi)^2} }\left.
                   - \int_{0}^{1}dx \; \frac{32}{3}g_{\pi N\Delta}^2 \Delta \left[\sqrt{
                   \Delta^2-m_{\pi}^2}\log R -\sqrt{\Delta^2-\tilde{m}^2}\log 
                   R[ \tilde{m}^2] \right] \right\}~, \\
F_{2}^{v}(q^2)&=& \kappa_v \left\{1- \frac{g_{A}^2}{\kappa_v} 
                  \frac{4\pi M_N}{(4\pi F_\pi)^2} 
                  \int_{0}^{1}dx\left[ \sqrt{\tilde{m}^2}-m_{\pi}\right]
                  \right. \nonumber \\
              & & \phantom{\kappa_v \;\;[1} 
                  +\frac{32g_{\pi N\Delta}^2 M_N \Delta}{9\kappa_v
                   (4\pi F_\pi)^2}\int_{0}^{1}dx \left[ \frac{1}{2}\log\left[
                   \frac{\tilde{m}^2}{4\Delta^2}\right]-\log\left[\frac{m_\pi}{
                   2\Delta}\right] \right. \nonumber \\
              & & \phantom{\kappa_v \;\;[1
                  +\frac{32g_{\pi N\Delta}^2 M_N \Delta}{9\kappa_v
                   (4\pi F_\pi)^2}\int_{0}^{1}dx \; } \left. \left. 
                  +\frac{\sqrt{\Delta^2-\tilde{m}^2}}{\Delta}
                  \log R\left[\tilde{m}^2\right]
                  -\frac{\sqrt{\Delta^2-m_{\pi}^2}}{
                  \Delta} \log R \right] \right\} \; .
\end{eqnarray}
To obtain a better understanding of the low energy structure of the nucleon 
as seen by electroweak probes, it is instructive to analyse the moments of
the form factors with respect to $q^2$:
\begin{eqnarray}
F_{i}^v(q^2)=F_{i}^v(0) \left[ 1 + \frac{1}{6} \; (r_{i}^v)^2 q^2
                               + {\cal O}(q^4)
                        \right]~.
\end{eqnarray}
The corresponding (squared) radii follow as
\begin{eqnarray}
\left(r_{1}^{v}\right)^2&=&6\frac{d F_{1}^{v}(q^2)}{d q^2}\biggl|_{q^2=0} 
                            \nonumber \\
                        &=&-\frac{1}{(4\pi F_\pi)^2}\left\{1+7 g_{A}^2 +
                           \left(10 g_{A}^2 +2\right) \log\left[\frac{m_\pi}{
                           \lambda}\right]\right\} -
                           \frac{12 B_{10}^{(r)}(\lambda)}{(4\pi F_\pi)^2} 
                           \nonumber \\
                        & &+\frac{g_{\pi N\Delta}^2}{54\pi^2 F_{\pi}^2}\left\{
                           26+30\log\left[\frac{m_\pi}{\lambda}\right]
                           +30\frac{\Delta}{\sqrt{\Delta^2-m_{\pi}^2}}
                           \log\left[\frac{\Delta}{m_\pi}+\sqrt{\frac{
                           \Delta^2}{m_{\pi}^2}-1}\right] \right\} \nonumber \\
                        &=& (0.67 +0.13)\,\, {\rm fm}^2 -
                           \frac{12 B_{10}^{(r)}(1\,{\rm GeV})}{(4\pi 
                           F_\pi)^2} 
                         = 0.80 \,\, {\rm fm}^2 -
                           \frac{12 B_{10}^{(r)}(1\,{\rm GeV})}{(4\pi F_\pi)^2} 
                           \; ,\\
\left(r_{2}^{v}\right)^2&=&\frac{6}{\kappa_v}\frac{d F_{2}^{v}(q^2)}{
                           d q^2}\biggl|_{q^2=0} \nonumber \\
                        &=&\frac{g_{A}^2 M_N}{8 F_{\pi}^2 \kappa_v \pi m_\pi}
                           +\frac{g_{\pi N\Delta}^2 M_N}{9 F_{\pi}^2 \kappa_v
                           \pi^2 \sqrt{\Delta^2-m_{\pi}^2}} \log\left[
                           \frac{\Delta}{m_\pi}+\sqrt{\frac{\Delta^2}{
                           m_{\pi}^2}-1}\right] \nonumber \\
                        &=&(0.52+0.09) \,\, {\rm fm}^2 = 0.61 \,\, {\rm fm}^2 
                           \; ,
\end{eqnarray}
Consider now the isovector Dirac radius. As it is well--known, it
diverges logarithmically in the chiral limit~\cite{BZ}. The most
precise empirical value is $(r_{1}^{v})^2 = 0.585\,$fm$^2$~\cite{MMD}.
Already the pion loop contribution slightly overshoots this value,
the $\Delta-\pi$ loop adds positively to this using the
renormalization scale of $\lambda
=1\,$GeV. In that case, we have to set $B_{10}^{(r)}(1\,{\rm GeV}) = 0.63$ to 
reproduce
the empirical value of $(r_{1}^{v})^2$. This number is of ``natural
size'' since we expect the LECs to be of order one in units of $1/16
\pi^2 F_\pi^2$.\footnote{To be precise, order one means any number between 0.1
  and 10. This range can be taken from the presently available
  determinations of LECs in the nucleon sector.} At present, we do not have 
a clear picture of the physics underlying $B_{10}^{(r)}$ that effectively 
shrinks the size of the pion-cloud in the Dirac form factor. In a 
``resonance-saturation'' model one would expect the $\rho$ vector meson to
play a large role in $B_{10}^{(r)}$. We also note that the strength of the
finite part of this counterterm can be brought to zero by reducing the
value of the renormalization scale $\lambda$ to 406 (858)~MeV for the small 
scale (chiral) expansion and simultaneously reproducing the empirical value of 
$(r_{1}^{v})^2$. Clearly more analysis is 
needed before the anatomy of this counterterm is understood. 

Much easier to interpret is
the isovector Pauli radius since to order $\epsilon^3$, it is free
of any LEC. Interestingly, the novel contribution from the
$\Delta-\pi$ loop leads to an increase of about 17\% and brings the
prediction  closer to the empirical
value, $(r_2^v)^2 = 0.80\,$fm$^2$. In the chiral limit, we recover the
well known $1/m_\pi$ singularity \cite{BZ}. In accord with the
decoupling theorem~\cite{gaze}, this singularity is not touched by the 
resonance contributions, i.e. the delta contribution vanishes for 
$\Delta \to \infty$.\footnote{It is amusing to note that for the
  special choice $\lambda =2\Delta$ the leading delta contribution
  also decouples for $r_1^v$. The relevance of this observation is,
  however, not  yet clear.}

In Fig.2, we show the $q^2$--dependence of the isovector form factors
in comparison to dispersion--theoretical result
of~\cite{MMD}. $F_1(q^2)$ is identical in the chiral and the small
scale expansion nicely reproducing the dispersion--theoretical result.
For $F_2(q^2)$, the $q^2$--dependence is  better reproduced in the
small scale than in the chiral expansion  although
the radius is still underestimated by about 15\%. This good description of
the isovector form factors is related to the two--pion continuum,
whose essential features on the left wing of the $\rho$--resonance are
reproduced in the one loop approximation. This is discussed in 
detail in~\cite{BKMFF}. In that paper, the chiral expansion of the
spectral functions was worked out to ${\cal O}(p^4)$ for the isovector
form factors and $\Delta (1232)$ effects were subsumed in the LEC $c_4$.
Finally, we compare the ${\cal O}(\epsilon^3)$ predictions for $F_{1}^v(q^2), 
F_{2}^v(q^2)$ of the ``small scale expansion'' with the corresponding 
ones of ${\cal O}(p^3)$ HBChPT. Switching off the delta degrees of freedom
via $g_{\pi N\Delta} \to 0$, the chiral
prediction for $F_{1}^v(q^2)$ is identical to the one shown in
Fig.~2, after readjusting the LEC ${B}_{10}$ to give the proper
radius. The resulting chiral prediction for $F_{2}^v(q^2)$ turns out to be 
worse than its ``small scale expansion'' analogue to this order 
(cf. the dot--dashed
lines in Fig.2), mostly due to the known underestimation of $(r_{2}^v)^2$
in HBChPT at ${\cal O}(p^3)$. Having discussed the isovector form factors
of the nucleon we now move on to the parallel discussion of the isoscalar
form factors.


\section{Isoscalar Vector Form factors}
\subsection{Definition}
In analogy to the isovector form factors one finds the Breit-frame reduction
of the isoscalar form factors:
\begin{eqnarray}
\langle N(p_2)|V_{\mu}^s(0)|N(p_1)\rangle=\frac{1}{N_1 N_2}
\bar{u}(r_2)\left[\tilde{G}_{E}^{\;s}(q^2)\;v_\mu
+\frac{1}{M_N}\tilde{G}_{M}^{\;s}(q^2)\left[
S_\mu,S_\nu\right] q^\nu\right]u(r_1) \times \eta^\dagger\frac{1}{2}\eta
\end{eqnarray}
We note here that the spectral functions of the isoscalar form factors
start at the three pion threshold and we thus can only have polynomial
terms at one loop order (compare e.g. \cite{BKMFF} for a more detailed
discussion). Again, for the choice $v_\mu = (1, \vec{0} \,)$, the 
$\tilde{G}_{E,M}^{\;s}$ correspond to the Sachs form factors $G_{E,M}^{\;s}$.
 
\subsection{Chiral Input}
The pertinent terms of the third order Lagrangian read:
\begin{eqnarray}
{\cal L}_{\pi N}^{(3)}&=& \frac{1}{(4\pi F_\pi)^2} \bar{N} \left\{
                          \tilde{B}_{1} D^\mu  v_{\mu\nu}^{(s)} v^\nu
                          + B_{20} \left({\rm Tr} (\chi_+) i v \cdot D
                          + {\rm h.c.}   \right)
                          + B_{29} \Delta i \left[S^\mu,S^\nu\right] 
                          v_{\mu\nu}^s 
                          + B_{30} \Delta^2 i v \cdot D \right\} N \nonumber \\
                      & & + \frac{1}{2 M_0} \; \bar{N} \left( \gamma_0 
                          B_{N}^{(2)\dagger}\gamma_0 B_{N}^{(1)} + \gamma_0 
                          B_{N}^{(1)\dagger}\gamma_0 B_{N}^{(2)}\right) N
                          - \frac{1}{(2 M_0)^2} \; \bar{N} \gamma_0 
                          B_{N}^{(1)\dagger}\gamma_0 \left(i v \cdot D
                          + \dot{g}_A S \cdot u \right) B_{N}^{(1)} N \; .
\end{eqnarray}
Here, $\tilde{B}_{1},B_{29}$ are {\em finite} ${\cal O}(\epsilon^3)$ 
counterterms for the reasons discussed above. In fact, $B_{29}$ gives
in principle a novel contribution to the isoscalar anomalous magnetic moment,
$\kappa_s =-0.12$. However, using again the decoupling argument, its
contribution has no physical consequence for the reason discussed
before, see also Table~2. As in the case of the
isovector form factors, we have a contribution from the nucleon $Z$
factor encoded in the LECs $B_{20}$ and $B_{30}$. Likewise, the pertinent
matrices $B_{N}^{(i)}$ are discussed in refs. \cite{BKKM,hhkbig}.

\subsection{Results}
In addition to the Born terms only diagrams 1b and 1f contribute to the
isoscalar vector form factors. We find
\begin{eqnarray}
G_{E}^{\;s}(q^2)&=&1+\kappa_s \frac{q^2}{4M_{N}^2}-4\tilde{B}_{1}
                   \frac{q^2}{(4\pi F_\pi)^2}\; ,  \\
G_{M}^{\;s}(q^2)&=&1+\kappa_s \; ,
\end{eqnarray}
with $\kappa_s$ given in Table 2. This leads to the Dirac and Pauli
form factors
\begin{eqnarray}
F_{1}^{s}(q^2)&=&1-4\tilde{B}_{1} \frac{q^2}{(4\pi F_\pi)^2} \; , \\
F_{2}^{s}(q^2)&=&\kappa_s \; .
\end{eqnarray}
The LEC $\tilde{B}_{1}$ can be determined from the empirical value
of $(r_1^s)^2 = (0.782\,{\rm fm})^2$,  $\tilde{B}_{1} = -0.88$. It is
again of ``natural size''. The physics underlying this counterterm has 
not been analysed yet in a systematic fashion, but from general 
phenomenological and symmetry considerations we expect  that a large
part of the finite value of this counterterm is related to the coupling
of the $\omega$ (and $\phi$) vector meson to the nucleon.  
All of the $q^2$-dependence of the isoscalar magnetic form factor 
and that of the isoscalar electric form factor beyond the radius
is therefore given by physics which is not accessible in a ${\cal O}(\epsilon^3)$ 
calculation, to be more precise, in a one--loop calculation. 
The corresponding spectral
functions have a cut starting at $t_0 = (3m_\pi)^2$ and thus only
give non--polynomial terms at two loop order (and higher).

\subsection{Results for the Sachs form factors}

{}From the isoscalar and isovector components, we can reconstruct the
proton and the neutron Sachs form factors,
\begin{equation}
G_{E,M}^p (q^2) = \frac{1}{2} \left(G_{E,M}^s (q^2) + G_{E,M}^v (q^2) 
\right)\,\, , \,\,\,
G_{E,M}^n (q^2) = \frac{1}{2} \left(G_{E,M}^s (q^2) - G_{E,M}^v (q^2) 
\right)\,\, . 
\end{equation}
In Fig.~3a, the resulting proton form factors are shown in comparison
to the dispersion--theoretical result~\cite{MMD} and the dipole fit 
\begin{equation}
G_{E}^p (q^2) = G_{M}^p (q^2) / \mu_p = G_{M}^n (q^2) / \mu_n 
= (1 - q^2/0.71\,{\rm GeV^2})^{-2}\,\, ,
\end{equation}
with $\mu_{p,n}$ the magnetic moment of the proton and the neutron, 
respectively. The corresponding neutron form factors are shown in Fig.~3b.
We note that since the isoscalar magnetic form factor is a constant to
this order, the $q^2$--dependence of $G_M^p$ and $G_M^n$ is identical.
This also holds for the chiral expansion to order $p^3$. The three
form factors $G_{E,M}^p$, $G_M^n$ are rather well described by the
$\epsilon^3$ results for momenta squared smaller than 0.2~GeV$^2$.  
The electric form factor $G_{E}^n(q^2)$ of the neutron is much smaller
in magnitude than  the other three form factors of the nucleon
(therefore, it is amplified in the figure by a factor of ten). 
The slope of $G_{E}^n(q^2)$ is related to the ``electric radius'' of the neutron via
\begin{eqnarray}
(r_{E}^n)^2&=& 6 \frac{d \; G_{E}^n(q^2)}{d q^2}\biggr|_{q^2=0} \nonumber \\ 
           &=& 6 \left[\frac{d \; F_{1}^n(q^2)}{d q^2}+\frac{\kappa_n}
            {4 M_{N}^2} \right]               \nonumber \\
           &=& - 0.113 \; {\rm fm}^2 \; ,
\end{eqnarray}
which is nearly completely dominated by the contribution from the anomalous magnetic 
moment (Foldy-term). The $q^2$--dependence of $G_E^n$ is reasonably
well described up to momenta squared of about 0.1~GeV$^2$. Note that
the smallness of this form factor stems from a cancellation of too
large contributions (isovector versus isoscalar currents) and thus
small inaccuarcies in describing either one of these show up visibly
in $G_E^n$.

\section{Axial Form Factors}

In addition to the electromagnetic form factors, the axial ones give further
information about the structure of the nucleon. In particular, the 
axial current is only partially conserved and the axial charge is not
equal to one. These facts are, of course, at the heart of the chiral
symmetry approach to the nucleon structure.

\subsection{Definition}
In the absence of second class currents, i.e. currents which have
opposite G--parity to the ones defined by the operators $\gamma_\mu
\gamma_5$ and $q_\mu \gamma_5$,\footnote{Experimentally, such second
  class currents are excluded to a high precision.} 
the most general matrix element of
the isovector axial current operator which conserves parity and time
reversal invariance is
\begin{eqnarray}
\langle N(p_2)|A_{\mu}^i(0)|N(p_1)\rangle=\bar{u}(p_2)\left[G_A (q^2)\;
\gamma_\mu\gamma_5+\frac{1}{2M_N}G_P(q^2)\;q_\mu \gamma_5
\right]u(p_1) \times \eta^\dagger\frac{\tau^i}{2}\eta~.
\end{eqnarray}
Here, $G_A (q^2)$ and $G_P (q^2)$ are the axial and the induced
pseudoscalar form factor, respectively. While $G_A (q^2)$ can be
extracted from (anti)neutrino--proton scattering or charged pion
electroproduction data, $G_P (q^2)$ can be determined in some
kinematical range from ordinary and radiative muon capture as well
as from pion electroproduction. 

We perform
a non-relativistic reduction of this matrix element in the Breit frame 
with the same kinematics used for the electromagnetic form factors. This gives
\begin{eqnarray}
\langle N(p_2)|A_{0}^i(0)|N(p_1)\rangle&=&0\\
\langle N(p_2)|A_{a}^i(0)|N(p_1)\rangle&=&\chi^{\dagger}_2 \left[\frac{E}{M_N}
                                          G_A (q^2)\; \sigma^a - \left( 
                                          \frac{E-M_N}{M_N}G_A (q^2)-
                                          \frac{q^2}{4 M_{N}^2}G_P(q^2)\right)
                                          \hat{q}^a \vec{\sigma} \cdot \hat{q}
                                          \right]\chi_1 \times \eta^\dagger
                                          \frac{\tau^i}{2}\eta \; ,
\end{eqnarray}
where $\chi$ denotes the 2-component spinor of the nucleon, $\sigma^a$
($a=1,2,3$) represents a Pauli matrix and $\hat{q}$ is a unit vector in the
direction of $\vec q\,$. The vanishing of the nucleon matrix element
$\langle N(p_2)|A_{0}^i(0)|N(p_1)\rangle$ is equivalent to the absence
of second class currents.

\subsection{Chiral Input}
First, we need some terms of the next--to--leading order meson
Lagrangian. Since our $\pi N$ Lagrangian is the heavy baryon version
of the form used in \cite{GSS}, we have to use the appropriate version
of ${\cal L}_{\pi\pi}^{(4)}$,
\begin{equation}
{\cal L}_{\pi\pi}^{(4)} = \frac{1}{16}l_3 {\rm
                          Tr}\left(\chi_+\right)^2+\frac{1}{16}
                          l_4\left[2 {\rm Tr}\left(\chi_+\right) 
                          {\rm Tr}\left(u\cdot u
                          \right)+2{\rm Tr}\left(\chi_{-}^2\right)-
                          {\rm Tr}\left(\chi_-\right)^2 \right] \,\, ,
\end{equation}
 with
\begin{eqnarray}
l_i&=&l_{i}^r(\lambda)+\gamma_i L ~. 
\end{eqnarray}
The LECs $l_3$ and $l_4$ contribute to the renormalization of the pion
mass and decay constant (using $\gamma_3 = 1/2, \gamma_4 =-2/3$),
\begin{eqnarray}
m_{\pi}^2 &=& m_{0}^2+\frac{m_0^4}{F_{\pi}^2}\left(2
  l_{3}^r(\lambda)+ \frac{1}{16\pi^2}\log\left[\frac{m_\pi}{\lambda}
  \right]\right)~, \\ 
F_\pi &=&{F}_0  + \frac{m_0^2}{F_0} \left(l_{4}^r(\lambda) -
\frac{1}{8\pi^2}\log\left[\frac{m_\pi}{\lambda}\right] 
  \right) \,\, ,
\end{eqnarray}
with $m_0^2 = (m_u+m_d)|\langle 0| \bar{q}q|0\rangle|/F_0^2$ the
leading term in the quark mass expansion of the pion mass squared and we
assume the standard scenario of dynamical chiral symmetry breaking
(large condensate). $F_0$ is the SU(2) chiral limit value of the pion
decay constant.
The relevant terms involving nucleons and deltas take  the form
\begin{eqnarray}
{\cal L}_{\pi N}^{(3)}&=&\frac{1}{(4\pi F_\pi)^2}\bar{N}\left\{
                         B_9 S\cdot u {\rm Tr}\left(\chi_+\right)
                         + \tilde{B}_{2} i \left[ S \cdot D , \chi_- \right]
                         + \tilde{B}_{3} S^\mu \left[ D^\nu , f_{\mu\nu}^-
                             \right]\right\}N \nonumber \\
                      & & - \frac{1}{(2 M_0)^2} \; \bar{N} \gamma_0 
                          B_{N}^{(1)\dagger}\gamma_0 \left(i v \cdot D
                          + \dot{g}_A S \cdot u \right) B_{N}^{(1)} N \; .
\end{eqnarray}
$\tilde{B}_{2},\tilde{B}_{3}$ are finite counterterms. Their meaning
is discussed in the next sections. In addition, the fixed $1/M^2$ 
contributions are solely governed by the matrix $B_{N}^{(1)}$, which can be 
found in refs. \cite{BKKM,hhkbig}.

\subsection{Results}
The chiral calculation yields results in the form
\begin{eqnarray}
\langle N(p_2)|A_{0}^i(0)|N(p_1)\rangle&=&0~,\\
\langle N(p_2)|A_{a}^i(0)|N(p_1)\rangle&=&\chi^{\dagger}_2 \left[
                                          G_1 (q^2)\; \sigma^a + G_2(q^2) \;
                                          \hat{q}^a \vec{\sigma} \cdot \hat{q}
                                          \right]\chi_1 \times \eta^\dagger
                                          \frac{\tau^i}{2}\eta \; .
\end{eqnarray}
Adding the loop amplitudes of Appendix B to the Born and counterterm 
contributions one finds in the Breit frame 
\begin{eqnarray}
G_{1}^{(3)}(q^2)&=&g_A - \frac{q^2}{8 M_{N}^2} g_A + \frac{q^2}{(4\pi 
                   F_\pi)^2} \tilde{B}_{3} \,\, ,\\
G_{2}^{(3)}(q^2)&=&\frac{q^2}{8 M_{N}^2}g_A-g_A \frac{q^2}{q^2-m_{\pi}^2}
                   +\frac{2 m_{\pi}^2 \tilde{B}_{2}}{(4\pi F_\pi)^2}\frac{
                   q^2}{q^2-m_{\pi}^2}-\tilde{B}_{3} \frac{q^2}{(4\pi 
                   F_\pi)^2} \,\, .
\end{eqnarray}
The rather lengthy renormalization of $g_A$ is given in Table~2.
However, these are not the quantities one is usually interested in.
The connection between $G_1(q^2), G_2(q^2)$ and the relativistic axial 
form factors $G_A(q^2), G_P(q^2)$ to ${\cal O}(\epsilon^3)$ in the
small scale expansion reads
\begin{eqnarray}
G_1(q^2)&=&G_A(q^2)-\frac{q^2}{8 M_{N}^2}g_A + {\cal O}(1/M_{N}^3) \; ,\\
G_2(q^2)&=&\frac{q^2}{8 M_{N}^2}g_A+\frac{q^2}{4 M_{N}^2} G_P(q^2) + 
           {\cal O}(1/M_{N}^3) \; ,
\end{eqnarray}
yielding
\begin{eqnarray}
G_A(q^2)&=&g_A + \frac{q^2}{(4\pi F_\pi)^2} \tilde{B}_{3} \; , \\
G_P(q^2)&=&\frac{4 M_{N}^2}{m_{\pi}^2-q^2}\left[g_A - \frac{2 m_{\pi}^2 
           \tilde{B}_{2}}{(4\pi F_\pi)^2}\right]-\tilde{B}_{3} \frac{4 
           M_{N}^2}{(4\pi F_\pi)^2} \; .
\end{eqnarray}
The meaning of the finite shift governed by $\tilde{B}_{2}$
together with the induced pseudoscalar form factor $G_P$ will be discussed
in the following section, whereas the finite counterterm $\tilde{B}_{3}$ 
has the simple interpretation as the radius of the axial form factor:
\begin{eqnarray}\label{eq:rA}
\left( r_A \right)^2&=& \frac{6}{g_A} \;\frac{d \; G_A(q^2)}{d q^2}\biggr|_{q^2=0} 
                        \nonumber \\
                    &=& \frac{\tilde{B}_{3}}{g_A}\;\frac{6}{(4\pi F_\pi)^2}
                        \equiv (0.65 \pm 0.03)^2 \,\, {\rm fm}^2\, ,
\end{eqnarray}
using the mean value of the axial radius deduced from
(anti)neutrino--proton scattering.\footnote{Here, we do not
consider the axial radius extracted from charged pion
electroproduction since a systematic analysis of such processes in the
small scale expansion is not yet available.}  This gives ${\tilde{B}_{3}} =
3.08 \pm 0.27$. Although we have fixed the numerical strength of this 
counterterm, more analysis is needed to understand the physics underlying
$\tilde{B}_3$. Presumably one will find an interplay between axial-vector 
meson $a_1$ and $\rho -\pi$ continuum contributions. Furthermore, as in the 
case of the isoscalar electromagnetic form factor, the
pertinent axial spectral function  starts with the three--pion
cut. Therefore, a one loop calculation can only lead to polynomial
contributions to the axial form factors. The small momentum behavior
of the axial spectral function in the framework of HBChPT is discussed
in \cite{BKMFF}. We note that $\Delta (1232)$ effects only appear in
the renormalization of the
axial--vector coupling constant (cf. Table~2) but not in the axial radius,
yielding exactly the same result for this form factor as in the chiral
$O(p^3)$ calculations \cite{BKKM,FLMS}, once the different coupling constant
renormalizations have been taken into account.

\section{Electroweak Form Factors and Muon Capture}

Here, we wish to discuss the induced pseudoscalar form factor and the
corresponding coupling  $g_P$ \cite{BKMGP,FLMS,ucebaf}. It is defined as the form factor for
the kinematics of ordinary muon capture on the proton (at rest),
\begin{eqnarray}
g_P&\equiv &\frac{m_\mu}{2 M_N} \; G_P(q^2=-0.88 m_{\mu}^2) \nonumber \\
   &=&\frac{2 m_\mu}{m_{\pi}^2+0.88 m_{\mu}^2}g_A M_N \left[1-\frac{2 m_{
      \pi}^2\tilde{B}_{2}}{(4\pi F_\pi)^2 g_A}\right]-\frac{1}{3}g_A m_\mu 
      M_N r_{A}^2 \nonumber \\
   &=&\frac{2 m_\mu F_\pi g_{\pi NN}}{m_{\pi}^2+0.88 m_{\mu}^2}-\frac{1}{3}
      g_A m_\mu M_N r_{A}^2 \nonumber \\
   &=& 8.46 \,\,(8.23) \label{eq:gp} \; ,
\end{eqnarray}
for $m_\mu = 0.106\,$GeV, $g_{\pi NN} = 13.4 \,\,(13.05)$
and we have made use of Eq.({\ref{eq:rA}) and of the definition 
of the Goldberger-Treiman deviation
\begin{equation}
\Delta_{\rm GT}\equiv  1-\frac{g_A M_N}{F_\pi g_{\pi NN}} 
           =-\frac{m_{\pi}^2 \tilde{B}_{2}}{8\pi^2 F_{\pi}^2
           g_A} \ll 1\,\,.
\end{equation}
The precise value of $\Delta_{\rm GT}$ depends on the values one
chooses for $g_A$ and $g_{\pi NN}$. In Table~3, taken from
\cite{ucebaf}, we have collected the most recent ranges allowed by
various input data.\footnote{Also given in that table is the
value of the monopole cut--off of the pion--nucleon vertex function, 
which can be extracted from the Goldberger--Treiman discrepancy under
the assumption that it is entirely due to the strong $\pi N$ form
factor.} We note that as in the case of $G_{A}(q^2)$, Eq.(\ref{eq:gp}), 
obtained in the small scale expansion including the $\Delta(1232)$, agrees 
exactly with the chiral prediction (obtained in HBChPT) to order $p^3$.
In fact, this remarkable prediction for $g_P$ agrees also with the pre-QCD
analysis of Adler and Dothan \cite{adler}\cite{wolf} and is extremely
precise \cite{BKMGP}. 

Accordingly, to third order both in HBChPT and in the small scale expansion 
the pseudoscalar form factor of the nucleon can be expressed as 
\begin{eqnarray}
G_{P}^{(3)}(q^2)&=&\frac{4 M_N g_{\pi NN} F_\pi}{m_{\pi}^2-q^2}-\frac{2}{3}
                   g_A M_{N}^2 r_{A}^2 \; , \label{eq:gpff}
\end{eqnarray}
which modifies the by now outdated pion pole (i.e. ${\cal O}(\epsilon^2)$)  
prediction
\begin{eqnarray}
G_{P}^{(2)}(q^2)&=&\frac{4 M_N g_{\pi NN} F_\pi}{m_{\pi}^2-q^2} \,\, .
\end{eqnarray}
Apart from the pion pole, one needs three--pion intermediate states
to build up additional non--polynomial dependence in $G_P (q^2)$ in the small
scale as well as the chiral expansion (i.e. one has to go to two loops).
So far, the only experiment in which the induced pseudoscalar form
factor was extracted in a certain kinematical range was the pion
electroproduction one at Saclay~\cite{saclay} (see also ~\cite{bkmsac}
for some comments). Within the precision of the experiment, the
$q^2$--dependence induced by the pion pole could be verified. More
precise data are certainly needed to further test the chiral/small
scale prediction for $G_P (q^2)$. 
Finally, we note that the recently published TRIUMF radiative muon capture 
experiment \cite{Triumf} finds a very large value for the induced 
pseudoscalar coupling constant, $g_P^{\rm Triumf} = (9.8\pm 0.7 \pm 0.3) \,
g_A$. This is about 50\% larger than the prediction in the chiral or
small scale expansion. It is not yet clear whether this result has to be considered
genuine or needs, e.g., to be subjected to further radiative (or other) 
corrections, see e.g.~\cite{FLMS}\cite{myhrer}\cite{min}\cite{fewbody}. 
{}From our calculation we must conclude  
that to this order $\Delta (1232)$ effects in the electroweak form factors 
can not explain the TRIUMF number. 

\section{Acknowledgments}

We would like to acknowledge helpful discussions with our colleague Hans-Werner
Hammer. This work was supported in part by the Natural Sciences and Engineering
Research Council of Canada.

\bigskip\bigskip 
\begin{appendix}
\section{Loop diagrams of the vector form factors}
There are six non-zero loop diagrams (cf. Fig.1) at ${\cal O}(\epsilon^3)$ 
contributing to the vector form factors. One finds
\begin{eqnarray}
A_{1a}&=&i \frac{g_{A}^2}{(4\pi F_{\pi})^2} \; \eta^\dagger \frac{\tau^i}{2} 
         \eta \left\{ \bar{u}(r_2) \; \epsilon_v \cdot v \; u(r_1) 
         \left[ \left(6 m_{\pi}^2 -\frac{5}{3}q^2\right)
         \left(16\pi^2 L + \log\frac{m_\pi}{\lambda}\right)+2 m_{\pi}^2
         -\frac{2}{3}q^2 \right. \right. \nonumber \\
      & &\phantom{i \frac{g_{A}^2}{(4\pi F_{\pi})^2} \; \eta^\dagger \frac{
         \tau^i}{2} \eta \; \{ \bar{u}(r_2) \; \epsilon_v \cdot v \; u(r_1) }
         \left. \left. +\int_{0}^{1}dx \left(3 m_{\pi}^2-5 q^2 x(1-x)\right)
         \log \left[\frac{\tilde{m}^2}{m_{\pi}^2}\right]
         \right] \right. \nonumber \\
      & &\phantom{i \frac{g_{A}^2}{(4\pi F_{\pi})^2} \; \eta^\dagger \frac{
         \tau^i}{2}
         \eta \;} \left. - \bar{u}(r_2) \left[ S_\mu , S_\nu \right]  
         \epsilon_{v}^\mu q^\nu \; u(r_1) \int_{0}^{1} dx \; 4\pi \; \sqrt{
         \tilde{m}^2} \right\} \\ 
A_{1b}&=&i \frac{g_{A}^2}{(4\pi F_{\pi})^2} \left\{ \eta^\dagger \frac{1}{2} 
         \; \eta \; \bar{u}(r_2) \; \epsilon_v \cdot v \; u(r_1) \left[ 
         \frac{3}{2} m_{\pi}^2 +\frac{9}{2} m_{\pi}^2 \left(16\pi^2 L+
         \log \frac{m_\pi}{\lambda}\right)\right] \right. \nonumber \\
      & &\phantom{i \frac{g_{A}^2}{(4\pi F_{\pi})^2} } \left. - \eta^\dagger 
         \frac{\tau^i}{2} \eta \; \bar{u}(r_2) \; \epsilon_v \cdot v \; u(r_1) 
         \left[ \frac{m_{\pi}^2}{2}+\frac{3}{2} m_{\pi}^2 \left(16\pi^2 L+
         \log \frac{m_\pi}{\lambda}\right)\right] \right\} \\
A_{1c}&=&i \frac{-1}{(4\pi F_{\pi})^2} \; \eta^\dagger \frac{\tau^i}{2} \eta
         \; \bar{u}(r_2) \; \epsilon_v \cdot v \; u(r_1) \left\{ 2 m_{\pi}^2
         \left(16\pi^2 L+\log\frac{m_\pi}{\lambda}\right)\right\} \\
A_{1d}&=&i \frac{1}{(4\pi F_{\pi})^2} \; \eta^\dagger \frac{\tau^i}{2} \eta
         \; \bar{u}(r_2) \; \epsilon_v \cdot v \; u(r_1) \left\{ \left(2
         m_{\pi}^2 -\frac{1}{3} q^2 \right) \left(16\pi^2 L+\log
         \frac{m_\pi}{\lambda}\right) + \int_{0}^{1} dx \; \tilde{m}^2
         \; \log \frac{\tilde{m}^2}{m_{\pi}^2} \right\} \\ 
A_{1e}&=&i \frac{8 g_{\pi N\Delta}^2}{(4\pi F_{\pi})^2} \; 
         \eta^\dagger \frac{\tau^i}{2} \eta
         \left\{ \bar{u}(r_2) \; \epsilon_v \cdot v \; u(r_1) 
         \left[ \left(\frac{2}{3} (2\Delta^2-m_{\pi}^2)+\frac{5}{27}q^2\right)
         \left(16\pi^2 L+\log
         \frac{m_\pi}{\lambda}\right)-\frac{1}{3}m_{\pi}^2+\frac{17}{162}q^2
         \right. \right. \nonumber \\
      & &\phantom{i \frac{8 g_{\pi N\Delta}^2}{(4\pi F_{\pi})^2} \; 
         \eta^\dagger \frac{\tau^i}{2} \eta \{ \bar{u}(r_2) \; 
         \epsilon_v \cdot v \; u(r_1) } 
         \; +\int_{0}^{1}dx \; \frac{1}{3}
         \left(2\Delta^2-m_{\pi}^2+\frac{5}{3}q^2 x(1-x)\right) 
         \log \frac{\tilde{m}^2}{m_{\pi}^2} \nonumber \\
      & &\phantom{i \frac{8 g_{\pi N\Delta}^2}{(4\pi F_{\pi})^2} \; 
         \eta^\dagger \frac{\tau^i}{2} \eta \{ \bar{u}(r_2) \; 
         \epsilon_v \cdot v \; u(r_1) } \left. \left.
         \; +\int_{0}^{1}dx \; \frac{1}{3} \left(
         \frac{4}{3}q^2 x(1-x)\frac{\Delta}{\sqrt{\Delta^2-\tilde{m}^2}}
         +4\Delta\sqrt{\Delta^2-\tilde{m}^2}\right) \log R[\tilde{m}^2] 
         \right] \right. \nonumber \\
      & &\phantom{i \frac{8 g_{\pi N\Delta}^2}{(4\pi F_{\pi})^2} \; 
         \eta^\dagger \frac{\tau^i}{2} \eta \; } + \bar{u}(r_2) 
         \left[ S_\mu , S_\nu \right] \epsilon_{v}^\mu q^\nu \; u(r_1) \left[ 
         \frac{4}{9}\Delta \left(
         16\pi^2 L +\log\frac{m_\pi}{\lambda}\right) -\frac{10}{27}\Delta
         \right. \nonumber \\
      & &\phantom{i \frac{8 g_{\pi N\Delta}^2}{(4\pi F_{\pi})^2} \; 
         \eta^\dagger \frac{\tau^i}{2} \eta \; + \bar{u}(r_2) 
         \left[ S_\mu , S_\nu \right] \epsilon_{v}^\mu q^\nu \; u(r_1) }
         \left. \left. +\int_{0}^{1}dx\left(\frac{2}{9}\Delta\log\left[
         \frac{\tilde{m}^2}{m_{\pi}^2}\right]
         +\frac{4}{9}\sqrt{\Delta^2-\tilde{m}^2}\log R[\tilde{m}^2]
         \right)\right]\right\} \\ 
A_{1f}&=&i \frac{-8 g_{\pi N\Delta}^2}{(4\pi F_{\pi})^2} \left\{ \eta^\dagger 
         \frac{1}{2} \; \eta \; \bar{u}(r_2) \; \epsilon_v \cdot v \; u(r_1) 
         \left[ (2\Delta^2-m_{\pi}^2)\left(16\pi^2 L +\log\frac{m_\pi}{\lambda}
         \right)-\frac{m_{\pi}^2}{2}+2\Delta\sqrt{\Delta^2-m_{\pi}^2}\log R 
         [m_{\pi}^2] 
         \right] \right. \nonumber \\
      & &\phantom{i \frac{-8 g_{\pi N\Delta}^2}{(4\pi F_{\pi})^2} } + 
         \eta^\dagger \frac{\tau^i}{2} \eta \; \bar{u}(r_2) \; \epsilon_v 
         \cdot v \; u(r_1) \left[ \frac{5}{3}(2\Delta^2-m_{\pi}^2)
         \left(16\pi^2 L 
         +\log\frac{m_\pi}{\lambda}\right)-\frac{5}{6}m_{\pi}^2 \right.
         \nonumber \\
      & &\phantom{i \frac{-8 g_{\pi N\Delta}^2}{(4\pi F_{\pi})^2} + 
         \eta^\dagger \frac{\tau^i}{2} \eta \; \bar{u}(r_2) \; \epsilon_v 
         \cdot v \; u(r_1) } \left. \left. +\frac{10}{3}
         \Delta\sqrt{\Delta^2-m_{\pi}^2}\log R [m_{\pi}^2] \right] 
         \right\} ,
\end{eqnarray}
with $\tilde{m}^2=m_{\pi}^2-q^2 x(1-x)$ and $R[z]=\frac{\Delta}{\sqrt{z}}+
\sqrt{\frac{\Delta^2}{z^2}-1}$.

\bigskip\bigskip

\section{Loop Diagrams of the Axial Form Factors}

There are 12 non-zero loop diagrams (cf. Fig.4) at ${\cal O}(\epsilon^3)$
contributing to the axial form factors. One finds
\begin{eqnarray}
A_{2a}&=&i \frac{g_A}{(4\pi F_\pi)^2} \;\eta^\dagger \frac{
         \tau^i}{2}\eta \;\bar{u}(r_2) \; \frac{S\cdot q \;\epsilon_a\cdot q}{
         q^2-m_{\pi}^2} \; u(r_1) \; 2 m_{\pi}^2 \left(16\pi^2 L +\log\frac{
         m_\pi}{\lambda} \right) \\
A_{2b}&=&i \frac{g_A}{(4\pi F_\pi)^2} \;\eta^\dagger \frac{
         \tau^i}{2}\eta \;\bar{u}(r_2) \; \frac{S\cdot q \;\epsilon_a\cdot q}{
         q^2-m_{\pi}^2} \; u(r_1) \left(16\pi^2 L +\log\frac{
         m_\pi}{\lambda} \right) \frac{4 m_{\pi}^2 q^2-6 m_{\pi}^4}{q^2-
         m_{\pi}^2} \\
A_{2c}&=&i \frac{- g_A}{(4\pi F_\pi)^2} \;\eta^\dagger \frac{
         \tau^i}{2}\eta \;\bar{u}(r_2) \; \frac{S\cdot q \;\epsilon_a\cdot q}{
         q^2-m_{\pi}^2} \; u(r_1) \; 2 m_{\pi}^2 \left(16\pi^2 L +\log\frac{
         m_\pi}{\lambda} \right) \\
A_{2d}&=&i \frac{- g_A}{(4\pi F_\pi)^2} \;\eta^\dagger \frac{
         \tau^i}{2}\eta \;\bar{u}(r_2) \; S\cdot \epsilon_a \; u(r_1) \;
         4 m_{\pi}^2 \left(16\pi^2 L +\log\frac{m_\pi}{\lambda} \right)\\
A_{2e}&=&i \frac{g_{A}^3}{(4\pi F_\pi)^2} \;\eta^\dagger \frac{
         \tau^i}{2}\eta \;\bar{u}(r_2) \; S\cdot \epsilon_a \; u(r_1) \;
         m_{\pi}^2 \left(16\pi^2 L +\log\frac{m_\pi}{\lambda} +1\right)\\
A_{2f}&=&i \frac{- g_{A}^3}{(4\pi F_\pi)^2} \;\eta^\dagger \frac{
         \tau^i}{2}\eta \;\bar{u}(r_2) \; \frac{S\cdot q \;\epsilon_a\cdot q}{
         q^2-m_{\pi}^2} \; u(r_1) \; m_{\pi}^2 \left(16\pi^2 L +\log\frac{
         m_\pi}{\lambda} + 1\right) \\
A_{2g}&=&i \frac{g_A g_{\pi N\Delta}^2}{(4\pi F_\pi)^2} \;\eta^\dagger \frac{
         \tau^i}{2}\eta \;\bar{u}(r_2) \; S\cdot \epsilon_a \; u(r_1) \;
         \frac{32}{9} \left\{ \left(2m_{\pi}^2-
         \frac{4}{3}\Delta^2\right)\left(16\pi^2 L +\log\frac{m_\pi}{\lambda}
         \right)+\frac{8}{9}\Delta^2-m_{\pi}^2 \right. \nonumber
         \\
      & &\phantom{i \frac{g_A g_{\pi N\Delta}^2}{(4\pi F_\pi)^2} \;
         \eta^\dagger \frac{\tau^i}{2}\eta \;\bar{u}(r_2) \; S\cdot \epsilon_a
         \; u(r_1) \; \frac{32}{9} } \left.
         +\frac{2}{3}\frac{\pi \; m_{\pi}^3}{\Delta}-\frac{4}{3}\frac{(
         \Delta^2-m_{\pi}^2)^{3/2}}{\Delta}\log R[m_{\pi}^2]\right\} \\
A_{2h}&=& A_{2g} \\
A_{2i}&=&i \frac{g_1 g_{\pi N\Delta}^2}{(4\pi F_\pi)^2} \;\eta^\dagger \frac{
         \tau^i}{2}\eta \;\bar{u}(r_2) \; S\cdot \epsilon_a \; u(r_1) \;
         \frac{400}{81} \left\{\left(2\Delta^2-m_{\pi}^2\right)\left(16\pi^2 L
         +\log\frac{m_\pi}{\lambda}\right)+\Delta^2-m_{\pi}^2+\frac{2\Delta^2-
         m_{\pi}^2}{30} \right. \nonumber \\
      & &\phantom{i \frac{g_1 g_{\pi N\Delta}^2}{(4\pi F_\pi)^2} \;
         \eta^\dagger \frac{\tau^i}{2}\eta \;\bar{u}(r_2) \; S\cdot 
         \epsilon_a \; u(r_1) \;\frac{400}{81} } \left.
         + 2 \Delta \sqrt{\Delta^2-m_{\pi}^2}\log R[m_{\pi}^2] \right\} \\  
A_{2j}&=&i \frac{- g_A g_{\pi N\Delta}^2}{(4\pi F_\pi)^2} \;\eta^\dagger \frac{
         \tau^i}{2}\eta \;\bar{u}(r_2) \; \frac{S\cdot q \;\epsilon_a\cdot q}{
         q^2-m_{\pi}^2} \; u(r_1) \; \frac{32}{9} \left\{ \left(2m_{\pi}^2-
         \frac{4}{3}\Delta^2\right)\left(16\pi^2 L +\log\frac{m_\pi}{\lambda}
         \right)+\frac{8}{9}\Delta^2-m_{\pi}^2 \right. \nonumber \\
      & &\phantom{i \frac{- g_A g_{\pi N\Delta}^2}{(4\pi F_\pi)^2} \;
         \eta^\dagger \frac{\tau^i}{2}\eta \;\bar{u}(r_2) \; \frac{S\cdot q 
         \;\epsilon_a\cdot q}{q^2-m_{\pi}^2}
         \; u(r_1) \; \frac{32}{9} } \left.
         +\frac{2}{3}\frac{\pi \; m_{\pi}^3}{\Delta}-\frac{4}{3}\frac{(
         \Delta^2-m_{\pi}^2)^{3/2}}{\Delta}\log R[m_{\pi}^2]\right\} \\
A_{2k}&=& A_{2j} \\
A_{2l}&=&i \frac{- g_1 g_{\pi N\Delta}^2}{(4\pi F_\pi)^2} \;\eta^\dagger \frac{
         \tau^i}{2}\eta \;\bar{u}(r_2) \; \frac{S\cdot q \;\epsilon_a\cdot q}{
         q^2-m_{\pi}^2} \; u(r_1) \;
         \frac{400}{81} \left\{\left(2\Delta^2-m_{\pi}^2\right)\left(16\pi^2 L
         +\log\frac{m_\pi}{\lambda}\right)+\Delta^2-m_{\pi}^2+\frac{2\Delta^2-
         m_{\pi}^2}{30} \right.\nonumber \\
      & &\phantom{i \frac{- g_1 g_{\pi N\Delta}^2}{(4\pi F_\pi)^2} \;
         \eta^\dagger \frac{\tau^i}{2}\eta \;\bar{u}(r_2) \; \frac{S\cdot q 
         \; \epsilon_a\cdot q}{q^2-m_{\pi}^2} \; u(r_1) \;\frac{400}{81} } 
         \left. + 2 \Delta \sqrt{\Delta^2-m_{\pi}^2}\log R[m_{\pi}^2] \right\}
\end{eqnarray}

\end{appendix}

\bigskip\bigskip


\vspace{2cm}
\vfill\eject

\centerline{ {\Large {\bf Tables}}}

\bigskip

\renewcommand{\arraystretch}{1.3}

\begin{table}
\begin{tabular}{|l|c|c|c|c|}
LEC & $O_i$ & $\beta_i$ at ${\cal O}(p^3)$ & $\beta_i$ at ${\cal O}(\epsilon^3)$ & phen. \\ \hline
$B_9$ &$S\cdot u \;Tr(\chi_+)$&$\frac{g_{A}}{2}-\frac{g_{A}^3}{8}$&$\frac{g_{A}}{2}-\frac{g_{A}^3}{8}-g_A g_{\pi N\Delta}^2 \frac{16}{9}+g_1 g_{\pi N\Delta}^2\frac{50}{81}$&$g_A$ \\ \hline
$B_{10}$ &$\left(D^\mu f_{\mu\nu}^+ v^\nu\right)$&$-\frac{1}{6}-\frac{5}{6}g_{A}^2$&$-\frac{1}{6}-\frac{5}{6}g_{A}^2+\frac{20}{27}g_{\pi N\Delta}^2$& $\left(r_{1}^v\right)^2$ \\ \hline
$B_{20}$ &Tr$(\chi_+)\;i v\cdot D+h.c.$&$-\frac{9}{16}\;g_{A}^2$&$-\frac{9}{16}\;g_{A}^2-\;g_{\pi N\Delta}^2$& $Z_N$ \\ \hline
$B_{28}$ &$i \left[S^\mu,S^\nu\right]f_{\mu\nu}^+$& - &$+\frac{8}{9}g_{\pi N\Delta}^2$& $\kappa_v$ \\ \hline
$B_{29}$ &$i \left[S^\mu,S^\nu\right]v_{\mu\nu}^s$& - & 0 & $\kappa_s$ \\ \hline
$B_{30}$ &$i v \cdot D$& - & $+16 g_{\pi N\Delta}^2$ & $Z_N$ \\ \hline
$B_{31}$ &$ S \cdot u$& - &$+g_A g_{\pi N\Delta}^2 \frac{128}{27}-g_1 g_{\pi N\Delta}^2 \frac{400}{81}$& $g_A$ \\
\end{tabular}
\bigskip
\caption{Comparison of the counterterm structures between HBChPT and the 
small scale expansion.}
\end{table}

\begin{table}
\begin{tabular}{|rcl|}
$\kappa_s$&=&$\dot{\kappa}_s - 4 B_{29} \frac{\Delta M_N}{(4\pi F_\pi)^2}$ \\ 
\hline
$\kappa_v$&=&$\dot{\kappa}_v - g_{A}^2 \frac{4\pi m_\pi M_N}{(4\pi F_\pi)^2}
-\frac{4\Delta M_N}{(4\pi F_\pi)^2}\left[B_{28}^{(r)}(\lambda)+\frac{20}{27}g_{\pi N\Delta}^2-\frac{8}{9}g_{\pi N\Delta}^2\log\frac{2\Delta}{\lambda}\right]+\frac{32}{9}g_{\pi N\Delta}^2\frac{M_N\Delta}{(4\pi F_\pi)^2}\left(\log\left[\frac{m_\pi}{2\Delta}\right]+\frac{\sqrt{\Delta^2-m_{\pi}^2}}{\Delta}\log R\right)$ \\ \hline
$g_A$&=&$\dot{g}_A Z_{N}^{(3)}+\frac{m_{\pi}^2}{32\pi^2F_{\pi}^2}\left[8 B_{9}^r(\lambda)+\log\left[\frac{m_\pi}{\lambda}\right]\left(g_{A}^3-4 g_A\right)+g_{A}^3\right]+\frac{\Delta^2}{(4\pi F_\pi)^2}B_{31}^r(\lambda)$ \\ 
 & &$\phantom{\dot{g}_A Z_{N}^{(3)}}+\frac{g_A g_{\pi N\Delta}^2}{32\pi^2F_{\pi}^2}\left(\frac{128}{9}m_{\pi}^2-\frac{256}{27}\Delta^2\right) \log\left[\frac{m_\pi}{\lambda}\right]+\frac{g_A g_{\pi N\Delta}^2}{32\pi^2F_{\pi}^2}\left(
\frac{512}{81}\Delta^2-\frac{64}{9}m_{\pi}^2+\frac{128}{27}\pi\frac{m_{\pi}^3}{\Delta}-\frac{256}{27}\frac{(\Delta^2-m_{\pi}^2)^{3/2}}{\Delta}\log R\right)$ \\ \
 & &$\phantom{\dot{g}_A Z_{N}^{(3)}}+\frac{g_1 g_{\pi N\Delta}^2}{32\pi^2F_{\pi}^2}\frac{400}{81}\left\{(2\Delta^2-m_{\pi}^2)\log\left[\frac{m_\pi}{\lambda}\right]+\Delta^2-m_{\pi}^2+\frac{2\Delta^2-m_{\pi}^2}{30}+2\Delta\sqrt{\Delta^2-m_{\pi}^2}\log R\right\}$ \\ 
\end{tabular}
\bigskip
\caption{Renormalization of coupling constants in the small scale 
expansion.}
\end{table}

\bigskip

\begin{table}[hbt]
\begin{tabular}{|c|c|c|c|c|c|}
    $g_A$&$F_\pi$ [MeV] &$g_{\pi N}$ & $\Delta_{\rm GT}$ [\%] & $\tilde{B}_{2}$ 
    & $\Lambda$ [MeV] \\
    \hline
    1.260 & 92.42 & 13.40 & 4.5 & -1.57 & 656 \\
    1.260 & 92.42 & 13.05 & 2.0 & -0.68 & 994 \\
    1.266 & 92.42 & 13.40 & 4.1 & -1.41 & 691 \\
    1.266 & 92.42 & 13.05 & 1.5 & -0.52 & 1135 \\
\end{tabular} 
\bigskip
\caption{The Goldberger--Treiman discrepancy $\Delta_{\rm GT}$ for
    various input parameters, as taken from [28].
    Also given are the LEC $\tilde{B}_{2}$ and the
    pion--nucleon cut--off as explained in the text. Note that for
    this table we have used more precise values of $F_\pi$ and $M_N$
    than elsewhere in the main text.}\label{tab:GTR} 
\end{table}


\vspace{1cm}

\centerline{ {\Large {\bf Figure Captions}}}

\begin{enumerate}
\item[Fig.1] One--loop diagrams contributing to the vector form factors. The
graphs a,b,c,d are the standard graphs (chiral expansion to order
$p^3$) whereas e and f
are the new contributions in the small scale expansion at ${\cal O}
(\epsilon^3)$. Nucleons, $\Delta$'s, pions and photons are denoted
by solid, double, dashed and wiggly lines, respectively.

\item[Fig.2] Small scale expansion of the isovector nucleon form
  factors (solid lines) in comparison to the empirical
  parameterization of \cite{MMD} (dashed lines) and the order $p^3$
  chiral expansion (dot--dashed lines). The lower (upper) set of curves
  refers to $F_1^v(q^2)$ ($F_2^v(q^2)$).

\item[Fig.3] (a) Proton Sachs form factors. Solid lines: Small scale
  expansion, dashed lines: dipole fit, dashed--dotted lines:
  dispersion--theoretical result~\cite{MMD}. The upper and the lower
  curves refer to $G_M^p (q^2)$ and  $G_E^p (q^2)$, respectively. 
  (b) Neutron Sachs
  form factors (the dipole prediction for $G_E^n$ is zero and not
  shown). The  upper and the lower
  curves refer to $G_E^n (q^2)$ and  $G_M^n (q^2)$, respectively. 
  Note that $G_E^n (q^2)$ is multiplied by a factor of ten.

\item[Fig.4]  One--loop diagrams contributing to the axial form
  factors (including the ones which simply renormalize the mass,
  the propagator and the decay constant of the pion). The
  graphs a,$\ldots$,f are the standard graphs  whereas h,$\ldots$,l
  are the new contributions in the small scale expansion at ${\cal O}
  (\epsilon^3)$. The circle--cross denotes the coupling to the external
  axial--vector source.

\end{enumerate}

\newpage

\newpage

\centerline{ {\Large {\bf Figures}}}

$\,$\vspace{2cm}
\begin{figure}[bht]


\centerline{ 
\epsfysize=2.5in 
\epsffile{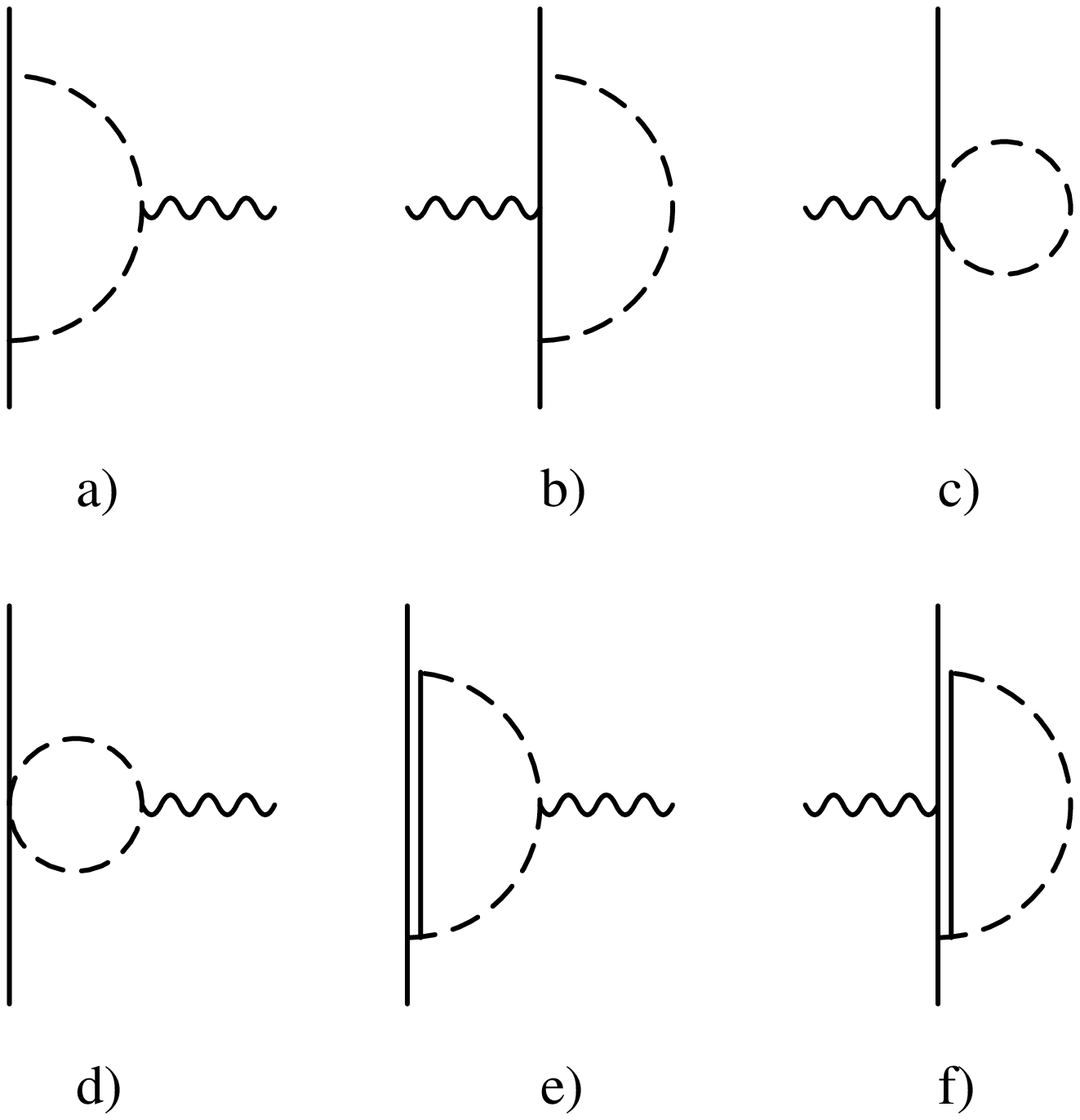}
}
\vskip 1cm 

\centerline{\Large Figure 1}
\end{figure}

\begin{figure}[bht]

$\;$\vspace{2cm}

\centerline{ 
\epsfysize=2.5in
\epsffile{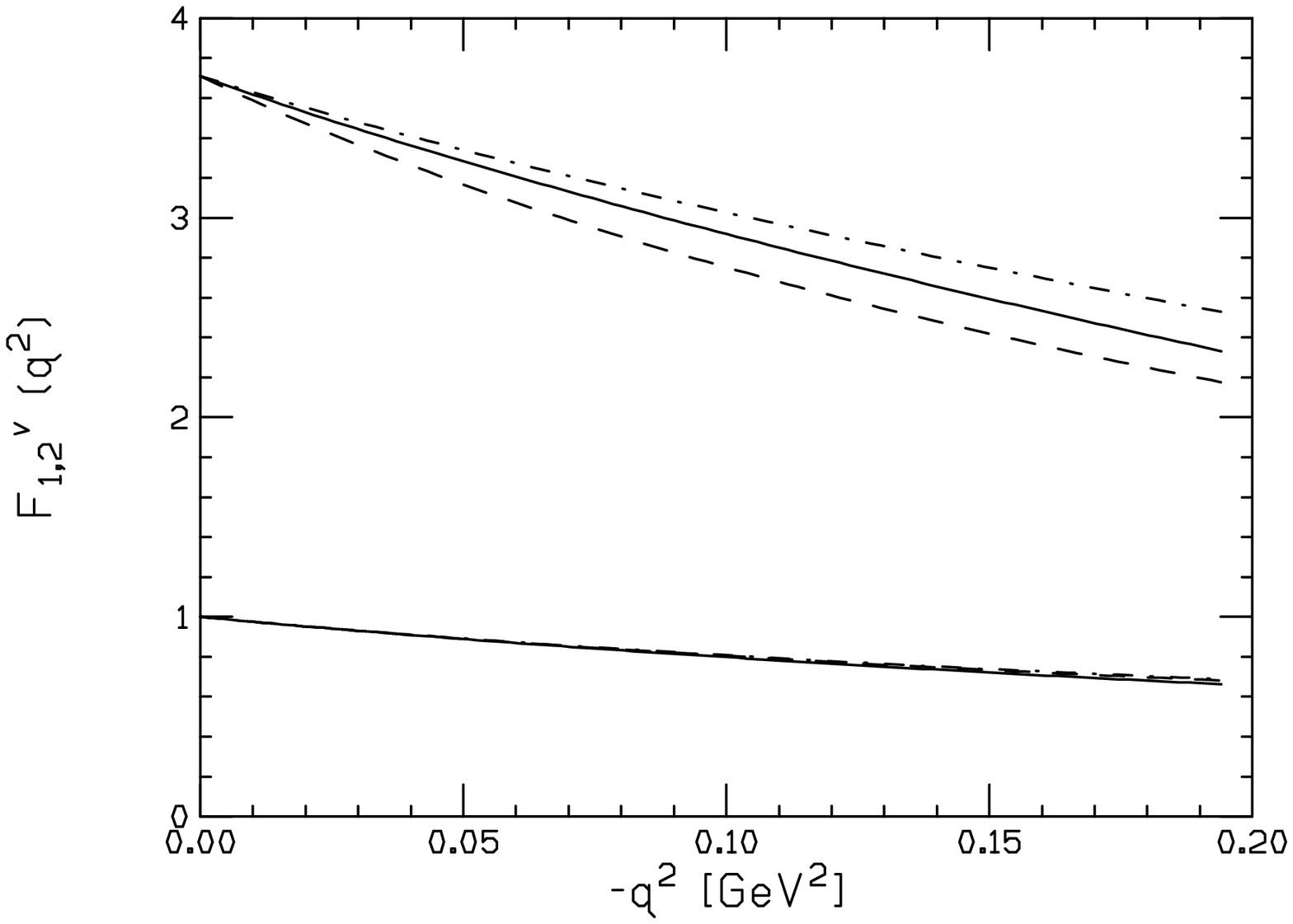}
}
\vskip 1cm

\centerline{\Large Figure 2}
\end{figure}

\newpage
\begin{figure}[bht]

$\;$\vspace{2cm}

\centerline{ 
\epsfysize=2.9in
\epsffile{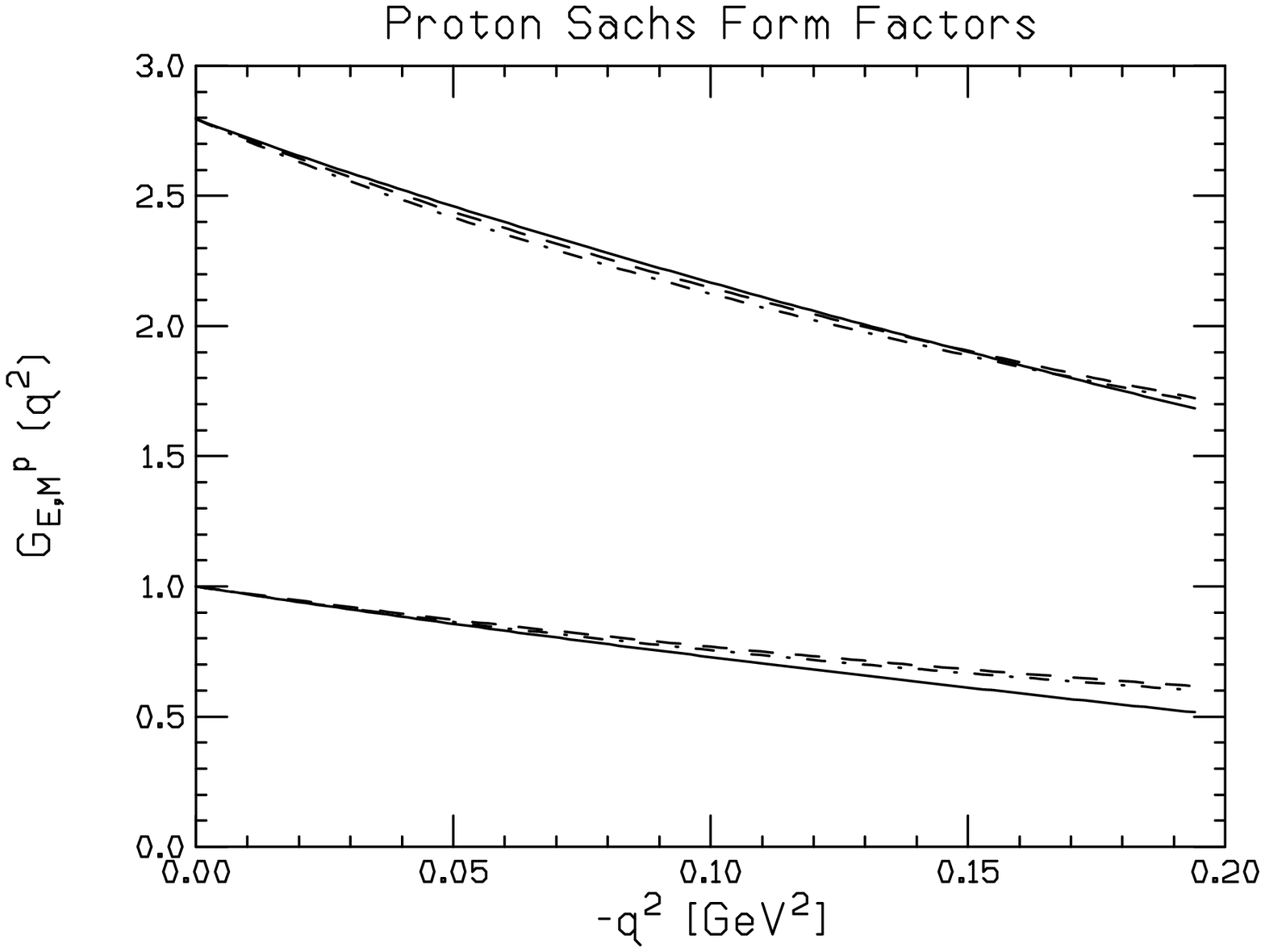}
}
\vskip 1cm

\centerline{\Large Figure 3a}
\end{figure}
\begin{figure}[bht]

$\;$\vspace{2cm}

\centerline{ 
\epsfysize=2.9in
\epsffile{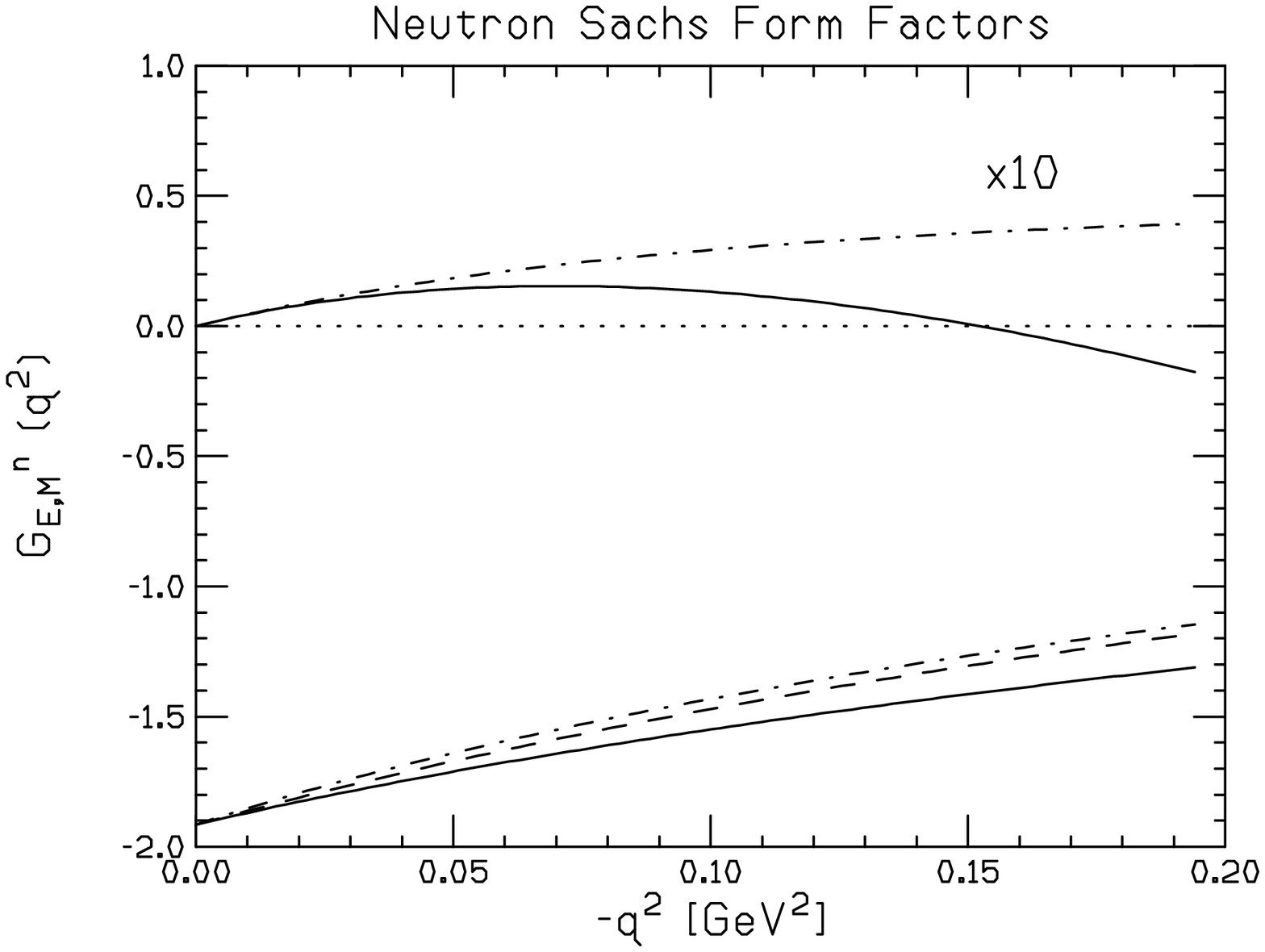}
}
\vskip 1cm

\centerline{\Large Figure 3b}
\end{figure}

\newpage

%
%
%
%
%

\begin{figure}[bht]

$\;$\vspace{2cm}

\centerline{ 
\epsfysize=3.5in
\epsffile{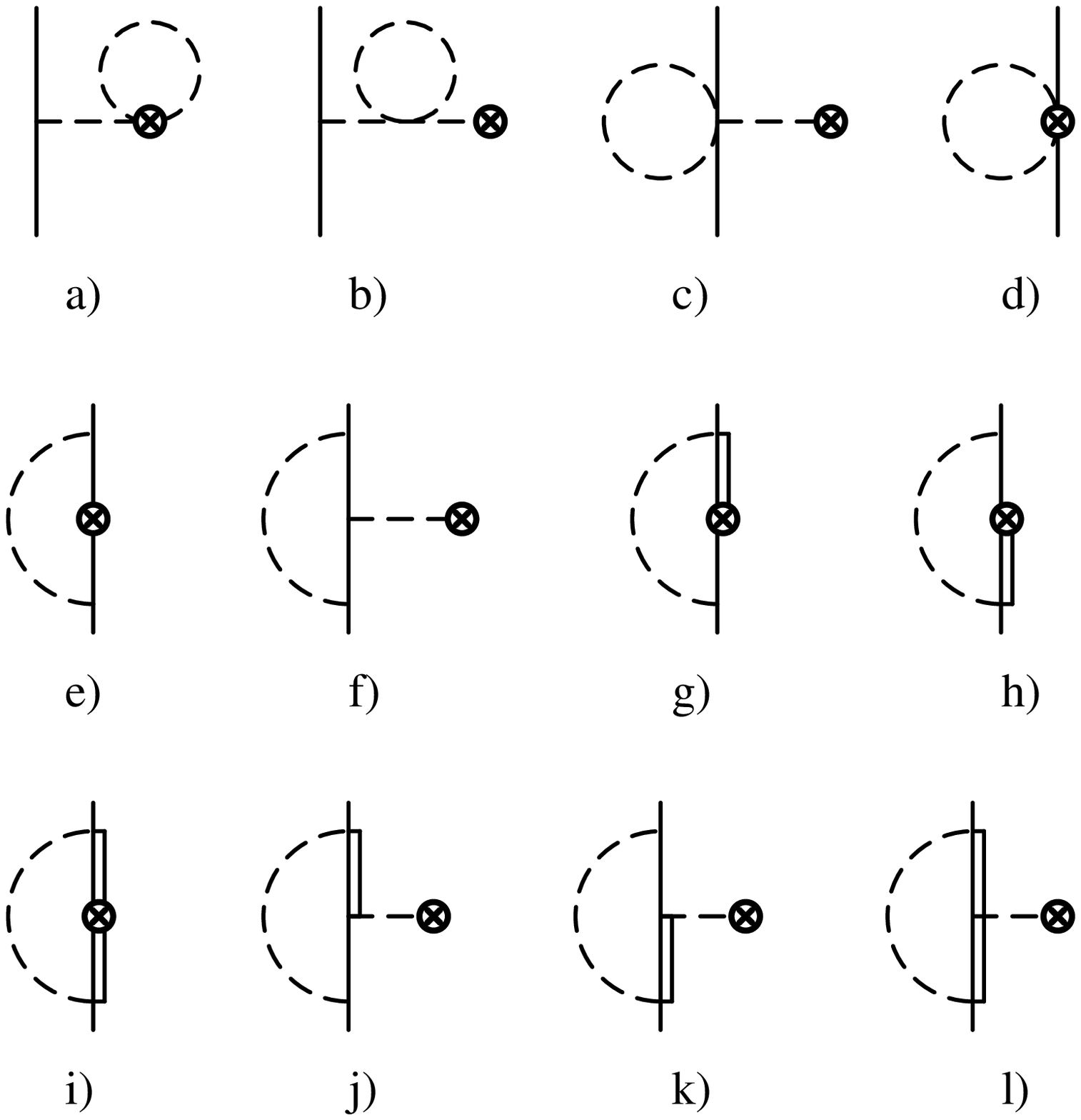}
}
\vskip 0.7cm

\centerline{\Large Figure 4}
\end{figure}

\end{document}